\documentclass[aps,prx,reprint,superscriptaddress,amsmath,amssymb]{revtex4-2}

\usepackage{bm}
\usepackage{braket}
\usepackage{graphicx}
\usepackage{txfonts}
\usepackage[hypertexnames=false]{hyperref}
\usepackage{xspace}

\newcommand{\secVC}{virial coefficient\xspace}
\newcommand{\secVCs}{virial coefficients\xspace}

\begin{document}

\title{Predicting heteropolymer interactions: demixing and hypermixing of disordered protein sequences}

\author{Kyosuke Adachi}
\affiliation{Nonequilibrium Physics of Living Matter RIKEN Hakubi Research Team, RIKEN Center for Biosystems Dynamics Research, 2-2-3 Minatojima-minamimachi, Chuo-ku, Kobe 650-0047, Japan}
\affiliation{RIKEN Interdisciplinary Theoretical and Mathematical Sciences Program, 2-1 Hirosawa, Wako 351-0198, Japan}
\author{Kyogo Kawaguchi}
\affiliation{Nonequilibrium Physics of Living Matter RIKEN Hakubi Research Team, RIKEN Center for Biosystems Dynamics Research, 2-2-3 Minatojima-minamimachi, Chuo-ku, Kobe 650-0047, Japan}
\affiliation{RIKEN Cluster for Pioneering Research, 2-2-3 Minatojima-minamimachi, Chuo-ku, Kobe 650-0047, Japan}
\affiliation{
Institute for Physics of Intelligence, The University of Tokyo, 7-3-1 Hongo, Tokyo 113-0033, Japan}
\affiliation{Universal Biology Institute, The University of Tokyo, Bunkyo-ku, Tokyo 113-0033, Japan}

\date{\today}

\begin{abstract}
Cells contain multiple condensates which spontaneously form due to the heterotypic interactions between their components.
Although the proteins and disordered region sequences that are responsible for condensate formation have been extensively studied, the rule of interactions between the components that allow demixing, i.e., the coexistence of multiple condensates, is yet to be elucidated.
Here we construct an effective theory of the interaction between heteropolymers by fitting it to the molecular dynamics simulation results obtained for more than 200 sequences sampled from the disordered regions of human proteins.
We find that the sum of amino acid pair interactions across two heteropolymers predicts the Boyle temperature qualitatively well, which can be quantitatively improved by the dimer pair approximation, where we incorporate the effect of neighboring amino acids in the sequences.
The improved theory, combined with the finding of a metric that captures the effective interaction strength between distinct sequences, allowed the selection of up to three disordered region sequences that demix with each other in multicomponent simulations, as well as the generation of artificial sequences that demix with a given sequence.
The theory points to a generic sequence design strategy to demix or hypermix thanks to the low dimensional nature of the space of the interactions that we identify.
As a consequence of the geometric arguments in the space of interactions, we find that the number of distinct sequences that can demix with each other is strongly constrained, irrespective of the choice of the coarse-grained model.
Altogether, we construct a theoretical basis for methods to estimate the effective interaction between heteropolymers, which can be utilized in predicting phase separation properties as well as rules of assignment in the localization and functions of disordered proteins.
\end{abstract}

\maketitle

\section{Introduction}

Proteins play a pivotal role in the spontaneous formation of membrane-less organelles within cells, driving the condensation process through a series of intricate molecular interactions~\cite{banani2017biomolecular,shin2017liquid,alberti2019considerations,sabari2020biomolecular,hirose2023guide,holehouse2023molecular}.  Intrinsically disordered regions (IDRs), characterized by the absence of canonical folded structures within protein sequences~\cite{van2014classification,uversky2017intrinsically}, have emerged as key mediators in orchestrating protein-protein and protein-RNA interactions, ultimately leading to the formation of these cellular condensates.
Interestingly, while these disordered regions lack well-defined structures, their sequence-specific interactions can rule subcellular localizations and organelle functionality. An outstanding question remains as to how the sequence-specific interactions, limited in diversity by the possible interactions between amino acid residues, can facilitate the coexistence of multiple phases within a cell.

Efforts have focused on understanding how the formation of condensates is driven by IDR in sequence-dependent manners. Researchers have identified specific segments within IDRs, commonly referred to as ``charge blocks'', characterized by regions enriched in either positively or negatively charged amino acids~~\cite{das2013conformations,nott2016membraneless}. Other studies have emphasized the significance of the overall amino acid composition, particularly the presence of charged and aromatic residues, in driving condensate formation rather than relying on precise sequence information~\cite{pak2016sequence}.
Regular spacing between specific amino acid residues, often termed ``stickers'' (e.g., aromatic residues)~\cite{lin2017intrinsically}, have been shown to play a role in facilitating multivalent yet weak interactions between proteins, which is crucial for the formation of condensates exhibiting liquid-like properties, distinguishing them from protein aggregates~\cite{wang2018molecular}.

A simplified model of IDR chains has been introduced as heteropolymers composed of monomers with bonds, where specific interactions are assigned between the monomers, representing amino acid residues, depending on their chemical properties.
In employing this scheme in molecular dynamics (MD) simulations, the hydrophobicity scale (HPS) model~\cite{dignon2018relation} uses the hydropathy parameter measured from the all-atom force field~\cite{kapcha2014simple}, or tuned to fit with \textit{in vitro} experiments on single-chain properties~\cite{tesei2021accurate}.
Another form of interaction has been introduced in the Mpipi model~\cite{joseph2021physics}, where the interactions involving aromatic residues have been designed to account for the role they play in IDR interactions~\cite{vernon2018pi}.
Due to the cost-effectiveness, these models have been utilized in large-scale simulations aiming to characterize the properties of IDRs in the proteome~\cite{tesei2024conformational, lotthammer2023direct} as well as in designing multiphase condensates by combining with genetic algorithms~\cite{chew2023thermodynamic,chew2023aromatic}.

While numerical simulations have provided valuable insights into IDR condensation, challenges remain in determining optimal interaction parameters, particularly considering the significant differences between \textit{in vitro} and \textit{in vivo} environments, due for example to the crowdedness and nonequilibriumness of the intracellular environment~\cite{hofling2013anomalous,harada2013reduced}.
A theoretical framework for IDR interactions that transcends specific models is essential to understanding selective condensation patterns observed in cells, represented by stress-responding bodies that only incorporate a certain set of molecules~\cite{kuechler2020distinct}, the coordination between transcription factors upon gene expression~\cite{sabari2018coactivator}, and the unmixing nature of membrane-less organelles~\cite{takakuwa2023shell}.

For an analytical approach toward estimating the interactions between polypeptide chains, the direct calculation of free energy in heteropolymer mixtures has been undertaken using the random phase approximation (RPA), with a primary focus on the charged residues and their long-range Coulomb interactions~\cite{lin2016sequence}. Another approach involves utilizing single-polymer properties to predict polymer-polymer interactions; studies have demonstrated correlations between the critical temperature and the Boyle temperature obtained from coarse-grained simulations, with single-molecule compactness~\cite{lin2017phase} as well as the theta temperature~\cite{dignon2018relation}. These findings highlight the utility of parameters derived from individual sequences, such as $\kappa$~\cite{das2013conformations}, sequence charge~\cite{sawle2015theoretical}, and hydropathy~\cite{zheng2020hydropathy} decorations.
Although there have been attempts to use these theories to predict and explain multiphase coexistence~\cite{lin2018theories,lin2017charge}, there remains a gap in the theoretical framework that can incorporate both charge and non-charge patterning to calculate interactions across distinct heteropolymers.
Considering the close-to-real interactions across residues will be crucial in understanding sequence-dependent selective condensation patterns observed in cells~\cite{greig2020arginine,lin2023dynamical} as well as in designing sequences or chemicals that will interact specifically with condensates~\cite{kamagata2019rational,kilgore2022learning}.

In this study, we take an empirical approach, employing MD simulations to construct and identify an analytical method that is useful in predicting interactions between heteropolymers. Our findings indicate that the effective interaction between polypeptides can be qualitatively estimated by considering the sum of interactions between monomers, with improvements achieved through the inclusion of neighboring pair contributions. We demonstrate the utility of this method in predicting demixing and hypermixing phenomena in multicomponent simulations and provide insights into generating multiple coexisting phases.

\section{Theory of IDR polymer interactions fit by simulation}
\label{Sec:prediction_boyle_critical}

A coarse-grained polymer model treats an IDR as a chain of $N$ amino acid monomers, which involves monomer-monomer interaction potentials $U_{a b} (r)$ that depend on the amino acid pair $\{ a, b \}$~\cite{dignon2018relation,tesei2021accurate,joseph2021physics}.
For the HPS model, we consider $U_{a b} (r) = U_{a b}^\mathrm{AH} (r) + U_{a b}^\mathrm{DH} (r)$, where $U_{a b}^\mathrm{AH} (r)$ is the pairwise potential of the Ashbaugh-Hatch form~\cite{ashbaugh2008natively}, and $U_{a b}^\mathrm{DH} (r)$ is the electrostatic potential with the Debye-H{\" u}ckel screening:
\begin{equation}
    U_{a b}^\mathrm{AH} (r) :=
    \left\{
    \begin{array}{l}
        U^\mathrm{LJ} (r; \epsilon, \sigma_{a b}) + \epsilon (1 - \lambda_{a b}) \ \ \ (r \leq 2^{1 / 6} \sigma_{a b}) \\
        \lambda_{a b} U^\mathrm{LJ} (r; \epsilon, \sigma_{a b}) \ \ \text{ (otherwise)},
    \end{array}
    \right.
\end{equation}
where $U^\mathrm{LJ} (r; \epsilon, \sigma) := 4 \epsilon [(\sigma / r)^{12} - (\sigma / r)^6]$ is the Lennard-Jones potential, and $\sigma_{a b} := (\sigma_a + \sigma_b) / 2$ and $\lambda_{a b} := (\lambda_a + \lambda_b) / 2$ are the average size and hydrophobicity scale of amino acid pair $\{ a, b \}$, respectively.
The electrostatic potential is defined as
\begin{equation}
    U_{a b}^\mathrm{DH} (r) := \frac{q_a q_b e^2}{4 \pi \varepsilon_0 \varepsilon_r r} e^{- r / D},
    \label{Eq:debye_huckel}
\end{equation}
where $q_a$ is the integer charge ($+ 1$ for $a \in \{ \mathrm{K}, \mathrm{R} \}$, $- 1$ for $a \in \{ \mathrm{D}, \mathrm{E} \}$, and $0$ otherwise), $e$ is the elementary charge, $\varepsilon_0$ is the vacuum permittivity, $\varepsilon_r$ is the relative permittivity, $D := \sqrt{\varepsilon_0 \varepsilon_r k_B T / (2 e^2 c_s)}$ is the Debye length, $k_B$ is the Boltzmann constant, $T$ is the temperature, and $c_s$ is the salt ionic strength.
It has been shown that tuning the parameters (i.e., $\epsilon$, $\{ \sigma_a \}$, and $\{ \lambda_a \}$) of this model can lead to predictions in simulations of real experiments~\cite{tesei2021accurate,tesei2024conformational}.
Hereafter, $k_B$ is set to unity.

To predict how two polymer chains will interact, we can calculate the quantity called the (second) \secVC,
\begin{equation}
    B (T) = 2 \pi \int _0 ^\infty dr \, r^2 [1 - e^{- U_\mathrm{PMF} (r) / T}],
    \label{Eq:excluded_volume}
\end{equation}
which reflects their overall attraction or repulsion.
Here, $U_\mathrm{PMF} (r)$ is the potential of mean force between two polymers, which is formally derived from $U_{a b} (r)$ as
\begin{equation}
    e^{- U_\mathrm{PMF} (r) / T} = \left \langle {e^{-\sum_{1 \leq n, m \leq N} U_{a_n a_m} (|\bm{r}_{1, n} - \bm{r}_{2, m}|) / T}} \right \rangle_r,
    \label{Eq:potential_of_mean_force}
\end{equation}
where $\bm{r}_{1, n}$ ($\bm{r}_{2, m}$) is the coordinate of the $n$th ($m$th) amino acid monomer that constitutes the first (second) polymer, and $a_n$ ($a_m$) is the corresponding amino acid type.
The canonical average $\braket{\cdots}_r$ is taken with a fixed distance ${r}$ between the centers of mass of the two polymers.

\begin{figure}[t]
    \centering
    \includegraphics[scale=1]{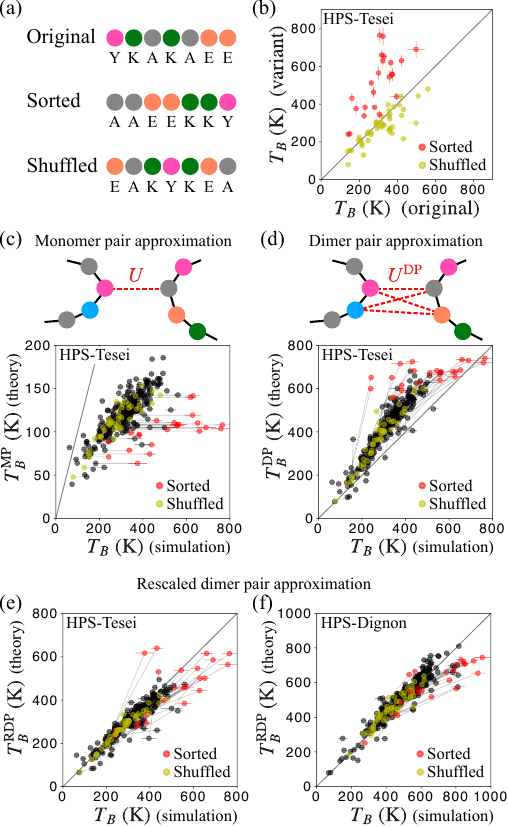}
    \caption{Prediction of the Boyle temperature.
    (a) Examples of the original, sorted, and shuffled polypeptide sequences.
    (b) The Boyle temperatures for the sorted (red) and shuffled (yellow) IDR sequences compared with the original sequences, obtained by simulations of the HPS-Tesei model.
    (c, d) The Boyle temperatures predicted by the (c) monomer pair or (d) dimer pair approximation, $T_B^\mathrm{MP}$ or $T_B^\mathrm{DP}$, respectively, for the original (black), sorted (red), and shuffled (yellow) sequences compared with the simulation result $T_B$ for the HPS-Tesei model.
    The thick gray line is $T_B^\mathrm{MP} = T_B$, and the thin gray lines connect the sorted or shuffled sequence to the corresponding original sequence.
    (e, f) The predicted Boyle temperatures $T_B^\mathrm{RDP}$ against the observed $T_B$ when using the rescaled dimer pair approximation with the rescaling parameter $\alpha = 0.75$ for the (e) HPS-Tesei and (f) HPS-Dignon models.}
    \label{Fig:boyle}
\end{figure}

\subsection{Boyle temperature predicted from amino acid composition}
\label{Sec:boyle_temperature}

The Boyle temperature $T_B$ is defined as the zero point of $B (T)$~\cite{landaubook}, which is the temperature where two polymer chains, on average, neither attract nor repel each other.
To understand how the specific types and order of amino acids influence the interactions between chains, we conducted MD simulations using the parameter set given in Ref.~\cite{tesei2021accurate} (HPS-Tesei) for over 50 different IDR sequences with $N=50$.
To seek an effective theory that predicts the phase separation properties of realistic protein sequences, we selected the disordered region sequences taken from the human proteome [Fig.~\ref{Fig:boyle}(a), see Appendix~\ref{App:selection_idr_sequences} for the selection of IDR sequences and the method of calculation].
We also generated some of their variants: alphabetically sorted and randomly shuffled sequences.
In Fig.~\ref{Fig:boyle}(b), we show how $T_B$ changes with sorting (red dots) or shuffling (yellow dots); sorting increases $T_B$ while shuffling typically decreases $T_B$, indicating that the effective interaction between the polymers is significantly different even with the same composition.

To begin explaining these simulation results, we use a simplified model that focuses on pairwise interactions between individual amino acids that appear when we expand the exponential factor in Eq.~\eqref{Eq:potential_of_mean_force} (Mayer $f$-function expansion, see Appendix~\ref{App:monomer_pair_approximation}):
\begin{equation}
    B^\mathrm{MP} (T) := \sum_{n = 1}^N \sum_{m = 1}^N v^\mathrm{MP}_{a_n a_m} (T).
    \label{Eq:excluded_volume_monomer_approx}
\end{equation}
Here, $n$ and $m$ are the monomer indices along the polymer, $N$ is the polymer length, and $v^\mathrm{MP}_{a b} (T)$ is the \secVC calculated at the monomer pair level:
\begin{equation}
    v^\mathrm{MP}_{a b} (T) := 2 \pi \int _0 ^\infty dr \, r^2 [1 - e^{- U_{a b} (r) / T}].
    \label{Eq:excluded_volume_monomer}
\end{equation}
To explicitly see how $B^\mathrm{MP} (T)$ depends on the amino acid composition, we can write 
\begin{equation}
    B^\mathrm{MP} (T) = \sum_{a, b} N_a N_b v^\mathrm{MP}_{a b} (T), 
    \label{Eq:excluded_volume_monomer_approx_amino_acid}
\end{equation}    
where $N_a$ is the total number of amino acids of type $a$ contained in a single IDR polymer.

This simplified value $B^\mathrm{MP} (T)$ is easier to calculate than the actual $B(T)$ because it ignores the details of how the polymer chain is shaped in 3D space.
Importantly, we can show that the approximation $B (T) \simeq B^\mathrm{MP} (T)$ becomes valid when taking the limit where the bond interaction within the polymer is very weak, or the bond length is much longer than the interaction range between the monomers (see Appendix~\ref{App:monomer_pair_approximation} for the detail).
$B^\mathrm{MP}$ has been used as one of the features to predict the properties of polypeptides~\cite{an2024active}.

To test how Eq.~\eqref{Eq:excluded_volume_monomer_approx} does well as an approximation, we compare $T_B$ obtained from simulations (see Appendix~\ref{App:simulation_excluded_volume}) and $T^\mathrm{MP}_B$ calculated as the zero point of $B^\mathrm{MP} (T)$ for over 250 different IDR sequences with $N=50$, including the sequences used in Fig.~\ref{Fig:boyle}(b) (see Appendix~\ref{App:selection_idr_sequences} for the selection of IDR sequences).
As plotted in Fig.~\ref{Fig:boyle}(c) with black dots, $T^\mathrm{MP}_B$ systematically underestimates $T_B$.
Nevertheless, we find a positive correlation between $T^\mathrm{MP}_B$ and $T_B$ with the Pearson correlation coefficient $r_P = 0.82$ and Spearman's rank correlation coefficient $r_S = 0.81$, suggesting that $B^\mathrm{MP} (T)$ qualitatively captures the composition dependence of $T_B$.

\begin{figure*}[t]
    \centering
    \includegraphics[scale=1]{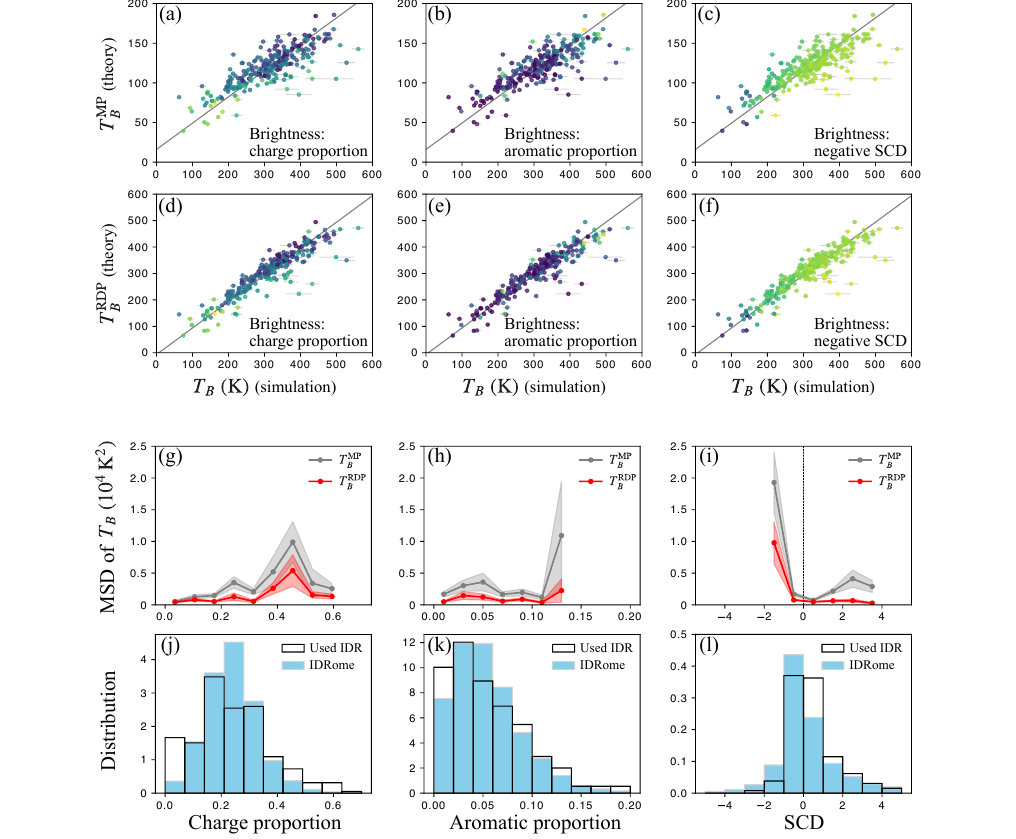}
    \caption{Feature dependence of the predicted Boyle temperature.
    (a-f) Boyle temperature predicted by the (a-c) monomer pair and (d-f) rescaled dimer pair approximations for the HPS-Tesei model [same as black dots in Figs.~\ref{Fig:boyle}(c) and (e), respectively], colored with the feature of each sequence: (a, d) charge proportion, (b, e) aromatic proportion, or (c, f) negative SCD.
    The gray line in each panel is the linear regression line.
    (g-i) MSD of $T_B$ measured from the regression line as a function of each feature with binning for the monomer pair (gray) and rescaled dimer pair (red) approximations.
    We plot the points with a count greater than $5$ in each bin.
    The shadow suggests the standard error.
    (j-l) Distribution of each feature for the IDR sequences used in simulations, compared with the distribution in the IDRome database~\cite{tesei2024conformational}.}
    \label{Fig:rev_boyle}
\end{figure*}

\subsection{Improved prediction using dimer pair approximation}
\label{Sec:improved_prediction}

The obvious shortcoming of $B^\mathrm{MP} (T)$ is that it only depends on the amino acid composition, and therefore would fail to capture the numerical and experimental results with chains with the same compositions but with distinct sequences.
To incorporate the sequence dependence in the calculation of the effective interactions, we consider the sum of the \secVCs for the pair of two neighboring amino acids (i.e., hetero-dimers), instead of the monomer pairs used in the simplest formula~\eqref{Eq:excluded_volume_monomer_approx}.
We introduce
\begin{equation}
    B^\mathrm{DP} (T) := \sum_{n = 1}^{N - 1} \sum_{m = 1}^{N - 1} v^\mathrm{DP}_{d_{n, n + 1} d_{m, m + 1}} (T),
    \label{Eq:excluded_volume_dimer_approx}
\end{equation}
where the \secVC for a dimer pair $\{ d_{n, n + 1}, d_{m, m + 1} \}$ is given as 
\begin{equation}
    v^\mathrm{DP}_{d_{n, n + 1} d_{m, m + 1}} (T) := 2 \pi \int _0 ^\infty dr \, r^2 \left[1 - e^{- U^\mathrm{DP}_{d_{n, n + 1} d_{m, m + 1}} (r) / T} \right],
    \label{Eq:excluded_volume_dimer}
\end{equation}
and the dimer-dimer interaction is defined as
\begin{equation}
    U^\mathrm{DP}_{d_{n, n + 1} d_{m, m + 1}} := U_{a_n a_m} + U_{a_n a_{m + 1}} + U_{a_{n + 1} a_m} + U_{a_{n + 1} a_{m + 1}}.
    \label{Eq:dimer_interaction_original}
\end{equation}

In essence, the dimer pair approximation [Eq.~\eqref{Eq:excluded_volume_dimer_approx}] is nothing but the monomer pair approximation [Eq.~\eqref{Eq:excluded_volume_monomer_approx}] except that the monomers are substituted with dimers. 
The intuition behind this procedure is that the effective bond interaction range will be enlarged (close to being doubled) while the interaction range is kept the same, which will bring the situation closer to the range where the monomer pair approximation is justified.
This approximation is distinct from calculating the next order of expansion of the exponential factor in the \secVC.
In fact, calculating the next order in the Mayer $f$-function expansion results in a useless approximation as its temperature dependence becomes non-monotonic (see Appendix~\ref{App:monomer_pair_zimm}).
An analogy can be found, however, in the calculation of annealing temperatures of nucleic acid sequences, where the interactions between the adjacent nucleotides are taken into account to improve the accuracy~\cite{santalucia1998unified}.

In Fig.~\ref{Fig:boyle}(d), we compare $T_B$ obtained from simulations with $T^\mathrm{DP}_B$ calculated as the zero point of $B^\mathrm{DP} (T)$.
Although $T^\mathrm{DP}_B$ systematically overestimates $T_B$ for the original sequences (black dots), we find a correlation between $T_B$ and $T^\mathrm{DP}_B$ with $r_P = 0.94$ and $r_S = 0.94$, which are higher than the result obtained from the monomer pair approximation.
Importantly, as shown with red and yellow dots in Fig.~\ref{Fig:boyle}(d), the current formula qualitatively captures the change upon ordering and shuffling the sequence, in contrast to Eq.~\eqref{Eq:excluded_volume_monomer_approx} [Fig.~\ref{Fig:boyle}(c)].

To fine-tune the accuracy of our model from Fig.~\ref{Fig:boyle}(d) by removing the systematic deviation, we introduce a scaling factor $\alpha$ that adjusts the strength of the interaction between dimer pairs.
In calculating the \secVC for a dimer pair $\{ d_{n, n + 1}, d_{m, m + 1} \}$, we use
\begin{equation}
    U^\mathrm{RDP}_{d_{n, n + 1} d_{m, m + 1}} := \alpha (U_{a_n a_m} + U_{a_n a_{m + 1}} + U_{a_{n + 1} a_m} + U_{a_{n + 1} a_{m + 1}}),
    \label{Eq:dimer_interaction_rescaled}
\end{equation}
instead of Eq.~\eqref{Eq:dimer_interaction_original}, and write the corresponding effective interaction and Boyle temperature as $B^\mathrm{RDP} (T)$ and $T^\mathrm{RDP}_B$, respectively.
Here, the parameter $\alpha$ can be interpreted as tuning the dimer-dimer interaction potential that is over-counted when calculating Eq.~\eqref{Eq:dimer_interaction_original}.
In Fig.~\ref{Fig:boyle}(e), we show the comparison of $T_B$ and $T^\mathrm{RDP}_B$ for $\alpha = 0.75$, which is approximately optimized to reduce the mean square relative error (MSRE) of $T^\mathrm{RDP}_B$ for the original sequences (see Appendix~\ref{App:rough_optimization_alpha} for the detail).
We see a quantitative agreement between $T_B$ and $T^\mathrm{RDP}_B$ ($\mathrm{MSRE} = 0.018$) while retaining a high correlation ($r_P = 0.93$ and $r_S = 0.92$).

To test whether this method of estimating $T_B$ is independent of the detail of the coarse-grained model, we changed the parameters of the amino acid size and hydrophobicity scale ($\{ \sigma_a \}$ and $\{ \lambda_a \}$) to values proposed in Ref.~\cite{dignon2018relation} (HPS-Dignon).
When we used a different set of parameters to describe the interactions between amino acids, the simulation results for $T_B$ changed considerably (Fig.~\ref{Fig:compare_boyle} in Appendix).
Nevertheless, $T_B$ can be correctly predicted by $T^\mathrm{RDP}_B$ calculated with the same $\alpha$ ($= 0.75$) [Fig.~\ref{Fig:boyle}(f)].
This result indicates that the composition of dimers is an important element in predicting Boyle temperatures for IDRs calculated by coarse-grained simulations.

We also find that the simulation results go off from the predictions by $T^\mathrm{RDP}_B$, especially for the sorted sequences.
This is likely due to the sorted sequences having longer regions of the same amino acids, represented by extended charged blocks.
To see this, we tested how the error in the prediction of $T_B$ depended on representative features: charge proportion, aromatic proportion, and sequence charge decoration (SCD)~\cite{sawle2015theoretical}, which quantifies the charge-blockiness.
In Figs.~\ref{Fig:rev_boyle}(a-f), we plot $T_B^\mathrm{MP}$ or $T_B^\mathrm{RDP}$ against the observed $T_B$ for the HPS-Tesei model, where the brightness indicates the value of each feature.
The gray line in each panel is the linear regression line, where we fitted $T_B = a T_B^\mathrm{MP(RDP)} + b$ with the parameters $a$ and $b$.
From Figs.~\ref{Fig:rev_boyle}(c) and (f), we see that negatively large SCD indicates higher $T_B$, which is captured by the monomer pair or the rescaled dimer pair approximation.
This tendency suggests that polymers with enhanced charge blocks (i.e., negatively large SCD) interact more attractively with each other~\cite{sawle2015theoretical}.

To examine whether the rescaled dimer pair approximation improves the prediction for sequences with a specific feature, we plot the mean square deviation (MSD) of $T_B$ from the linear regression line against the value of each feature [Figs.~\ref{Fig:rev_boyle}(g-i)].
We find that $T_B^\mathrm{RDP}$ (red lines) is typically improved from $T_B^\mathrm{MP}$ (gray lines) regardless of the feature type, suggesting that the rescaled dimer pair approximation is useful for a broad spectrum of IDR sequences.
As shown in Figs~\ref{Fig:rev_boyle}(j-l), the feature distribution for the sequences used in this study is similar to that for a large IDR database (IDRome~\cite{tesei2024conformational}), indicating that within the distribution of natural IDR sequences, the rescaled dimer pair approximation gives a reliable prediction of $T_B$. The accurate prediction of the property of sequences, including those with a high fraction of charged/aromatic residues or extended charged blocks, would likely require taking into account the longer-range interactions beyond the dimer pair approximation. Nevertheless, we continue here with the dimer pair approximation as it captures the sequence-dependent property of a majority of IDR sequences pooled from the human proteome.

\begin{figure}[t]
    \centering
    \includegraphics[scale=1]{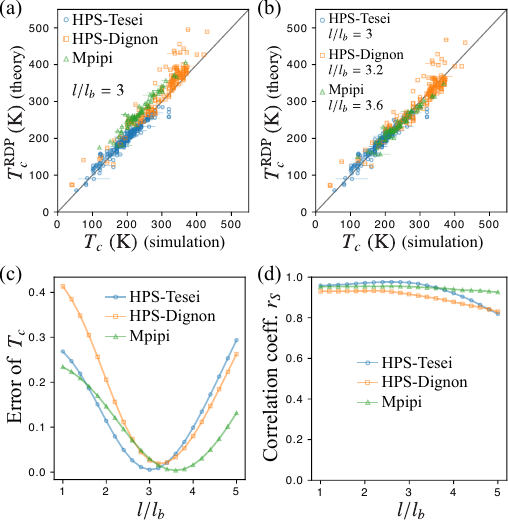}
    \caption{Prediction of the critical temperature for phase separation.
    (a,~b) Predicted critical temperatures $T_c^\mathrm{RDP}$ against $T_c$ observed in simulations for the HPS-Tesei (blue circles), HPS-Dignon (orange squares), and Mpipi (green triangles) models.
    The gray line is $T_c^\mathrm{RDP} = T_c$.
    In (a), we used the lattice constant parameter $l / l_b = 3$ for all the models.
    In (b), we took $l / l_b = 3$, $3.2$, and $3.6$ for the HPS-Tesei, HPS-Dignon, and Mpipi models, respectively, so that the error of $T_c$ is approximately minimized (see Appendix~\ref{App:rough_optimization_l} for the definition of the error).
    (c) Error of $T_c$ against $l / l_b$ for each model.
    (d) Spearman's rank correlation coefficient $r_S$ against  $l / l_b$ for each model.}
    \label{Fig:critical}
\end{figure}

\subsection{Critical temperature estimated from two-body interaction}
\label{Subsec:critical_temperature}

In the simulations of IDR polymers~\cite{dignon2018relation}, the Boyle temperature $T_B$ has been found to be highly correlated with the critical temperature for phase separation, $T_c$, which suggests that $T_c$ can be estimated from the effective two-body interaction $B (T)$.
To connect $B (T)$ to $T_c$ for IDR polymers, we consider the Flory-Huggins (FH) free energy~\cite{rubinstein2003polymer}: 
\begin{equation}
F = T \Omega [N^{-1} \phi \ln \phi + (1 - \phi) \ln (1 - \phi) + \chi (T) \phi^2].
\end{equation}
Here, $\Omega$ is the total number of sites in the lattice setup, $N$ is the polymer length, $\phi$ is the volume fraction of polymers, and $\chi (T)$ represents dimensionless interaction strength between polymers.

Since the FH theory considers polymers in a lattice system, the correspondence to the lattice-free polymer models requires further specification.
Assuming the correspondence in the dilute regime where the two-body interaction is dominant, we obtain  $\chi (T) = B (T) / (N^2 l^3) - 1 / 2$.
Here, $l$ is the lattice constant that needs to be specified in comparing with the FH theory, which we here treat as a fitting parameter that should be on the order of the bond length in the simulated IDR polymers, $l_b := 0.38\text{ nm}$.
We further assume that $\chi (T)$ has the form $\chi (T) = A_0 - 1 / 2 - B_0 / T$, where we subtract $1 / 2$ for convenience, which is often used~\cite{rubinstein2003polymer} and has been verified in IDR simulations~\cite{dignon2018sequence}.
This form can be derived when we assume that the exclusive part of the interaction potential is diverging and the attractive part is small~\cite{rubinstein2003polymer,landaubook}.

Under this setup, we obtain 
\begin{equation}
B (T) = N^2 l^3 \left(A_0 - \frac{B_0}{T} \right),
\end{equation}
which leads to $T_B = B_0 / A_0$.
According to the FH theory~\cite{rubinstein2003polymer}, the critical temperature for phase separation, $T_c$, is obtained as $T_c = B_0 / [A_0 + N^{- 1/ 2} + (2 N)^{-1}]$.
Thus, fitting the functional form of $B (T)$ around $T_B$ for a given $l$, we can find the optimal $A_0$ and $B_0$, from which we obtain the value of $T_c$.
Following this procedure, we obtain the estimated critical temperature $T^\mathrm{RDP}_c$ by assuming $B (T) \simeq B^\mathrm{RDP} (T)$ with $\alpha=0.75$.

To compare $T^\mathrm{RDP}_c$ with $T_c$ from numerics, we run phase separation simulations of IDR polymers using the interaction functions and parameters from Ref.~\cite{tesei2021accurate} (HPS-Tesei) (see Appendix~\ref{App:simulation_critical_temperature} for the detail of the simulation).
In Fig.~\ref{Fig:critical}(a) (with blue circles), we compare $T_c$ obtained from numerical simulations and $T^\mathrm{RDP}_c$ for a similar set of IDRs as used in Figs.~\ref{Fig:boyle}(c-e), and find that the correlation is high ($r_P = 0.97$ and $r_S = 0.97$).
Here, we selected $l$ as $l = 3 l_b$ by fitting so that the error between $T_c$ and $T^\mathrm{RDP}_c$ is approximately minimized, as plotted in Fig.~\ref{Fig:critical}(c) ($\mathrm{MSRE} = 0.0052$, see Appendix~\ref{App:rough_optimization_l} for the detail).
Further, as shown with orange squares in Fig.~\ref{Fig:critical}(a), we find that $T_c$ and $T^\mathrm{RDP}_c$ obtained for distinct model parameters~\cite{dignon2018relation} (HPS-Dignon) are highly correlated ($r_P = 0.95$ and $r_S = 0.92$) when using the same $\alpha = 0.75$ and $l= 3 l_b$.

Using the same procedure, we compare $T_c$ and $T^\mathrm{RDP}_c$ from the Mpipi model introduced in Ref.~\cite{joseph2021physics}, which has the residue-level interaction potential $U_{a b} (r) = U^\mathrm{WF}_{a b} (r) + U^\mathrm{DH}_{a b} (r)$, where the pairwise potential $U^\mathrm{WF}_{a b} (r)$ is a type of Wang-Frenkel potential defined by
\begin{equation}
    U^\mathrm{WF}_{a b} (r) :=
    \left\{
    \begin{array}{l}
         \tilde{\epsilon}_{a b} [(\tilde{\sigma}_{a b} / r)^{2 \mu_{a b}} - 1] [(3 \tilde{\sigma}_{a b} / r)^{2 \mu_{a b}} - 1]^2 \ \ \ (r \leq 3 \tilde{\sigma}_{a b}) \\
         0 \ \ \text{ (otherwise)}.
    \end{array}
    \right.
\end{equation}
We use the parameter values for $\{ \tilde{\epsilon}_{a b} \}$, $\{ \tilde{\sigma}_{a b} \}$, and $\{ \mu_{a b} \}$ proposed in Ref.~\cite{joseph2021physics}.
The electrostatic potential $U^\mathrm{DH}_{a b} (r)$ is given by Eq.~\eqref{Eq:debye_huckel} as before but with different values of $q_a$ ($0.75$ for $a \in \{ \mathrm{K}, \mathrm{R} \}$, $-0.75$ for $a \in \{ \mathrm{D}, \mathrm{E} \}$, $0.375$ for $a = \mathrm{H}$, and $0$ otherwise) and fixed values of the Debye length and relative permittivity ($D = 0.795 \text{ nm}$ and $\varepsilon_r = 80$).
This model has been shown to quantitatively reproduce the experimental result of the change in $T_c$ for several variants of the low complexity domain of heterogeneous nuclear ribonucleoprotein A1 (hnRNPA1)~\cite{joseph2021physics}.

The comparison between the prediction and numerical experiment is plotted with green triangles in Fig.~\ref{Fig:critical}(a), showing a high correlation ($r_P = 0.97$ and $r_S = 0.95$) and a good agreement even though the potential form is different from the HPS model.
Overall, the obtained results show that $B^\mathrm{RDP} (T)$, using two parameters obtained from a single model fitting ($\alpha=0.75$ and $l = 3 l_b$), is practically useful in estimating $T_c$ and $T_B$ even under the switching of residue-level interaction rules.

For the HPS-Dignon and Mpipi models, we can tune the parameter $l / l_b$ to minimize the error of $T_c$ [Fig.~\ref{Fig:critical}(c)], instead of using the same $l = 3 l_b$ as optimized for the HPS-Tesei model.
In Fig.~\ref{Fig:critical}(b), with the approximately optimized $l/l_b$ for each model, we plot the predicted $T_c^\mathrm{RDP}$ against $T_c$ from simulations, which suggests a better fit than the prediction with $l/l_b=3$ [Fig.~\ref{Fig:critical}(a)].
The appropriate lattice constant parameter $l$ is likely determined by the details of the interaction potential; the Wang-Frenkel potential in the Mpipi model has a slightly longer range compared with the HPS models, although it is not enough to explain the difference (the simple mean of $\tilde{\sigma}$ is $0.614$ in Mpipi whereas the mean of $\sigma$ is $0.588$ in HPS).
Nevertheless, the high correlation was maintained for a wide range of $l/l_b$ [Fig.~\ref{Fig:critical}(d)].

\begin{figure*}[t]
    \centering
    \includegraphics[scale=1]{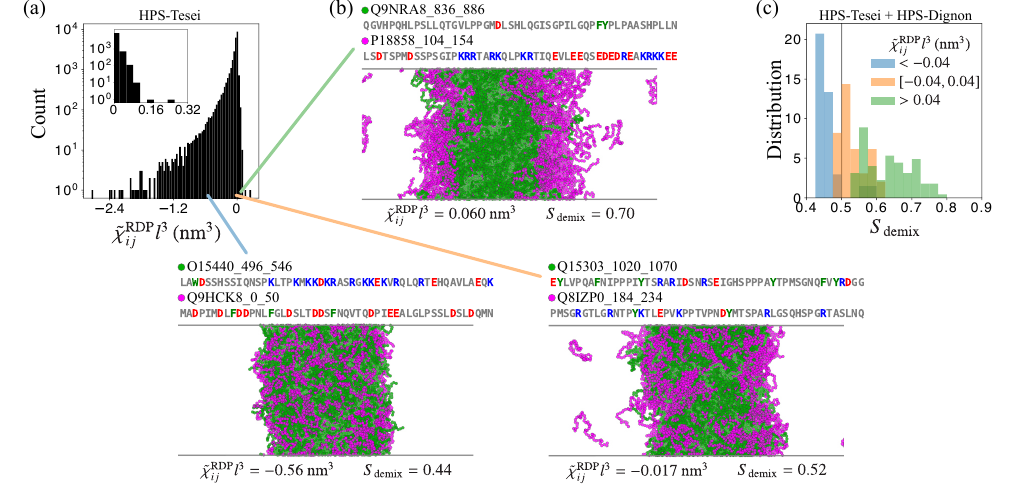}
    \caption{Prediction of hypermixing and demixing for two-component IDR polymers.
    (a) The histogram of the effective inter-component interaction parameter, $\tilde{\chi}_{ij}^\mathrm{RDP}$, at $T = \min \{ T_{c, i}, T_{c, j} \}$.
    The inset is the enlarged view of the positive region of $\tilde{\chi}_{ij}^\mathrm{RDP}$.
    (b) Snapshots of the simulated two-component IDR sequences (green and magenta), which are sampled from the three regions of $\tilde{\chi}_{ij}^\mathrm{RDP}$: $\tilde{\chi}_{ij}^\mathrm{RDP}l^3 < -0.04 \ \mathrm{nm}^3$ (bottom left), $-0.04 \ \mathrm{nm}^3 < \tilde{\chi}_{ij}^\mathrm{RDP}l^3 < 0.04 \ \mathrm{nm}^3$ (bottom right), and $\tilde{\chi}_{ij}^\mathrm{RDP}l^3 > 0.04 \ \mathrm{nm}^3$ (top).
    The sequences with their names are shown above each configuration.
    The bottom left and top configurations show hypermixing (with $S_\mathrm{demix} < 0.5$) and demixing (with $S_\mathrm{demix} > 0.5$), respectively.
    (c) The distribution of the demixing score $S_\mathrm{demix}$ for the three regions of $\tilde{\chi}_{ij}^\mathrm{RDP}$.
    The gray line suggests $S_\mathrm{demix} = 0.5$, which should be realized for a random mixture of the two components.
    The regions with $S_\mathrm{demix} < 0.5$ and $S_\mathrm{demix} > 0.5$ indicate hypermixing and demixing, respectively.}
    \label{Fig:demixing}
\end{figure*}

\begin{figure*}[t]
    \centering
    \includegraphics[scale=1]{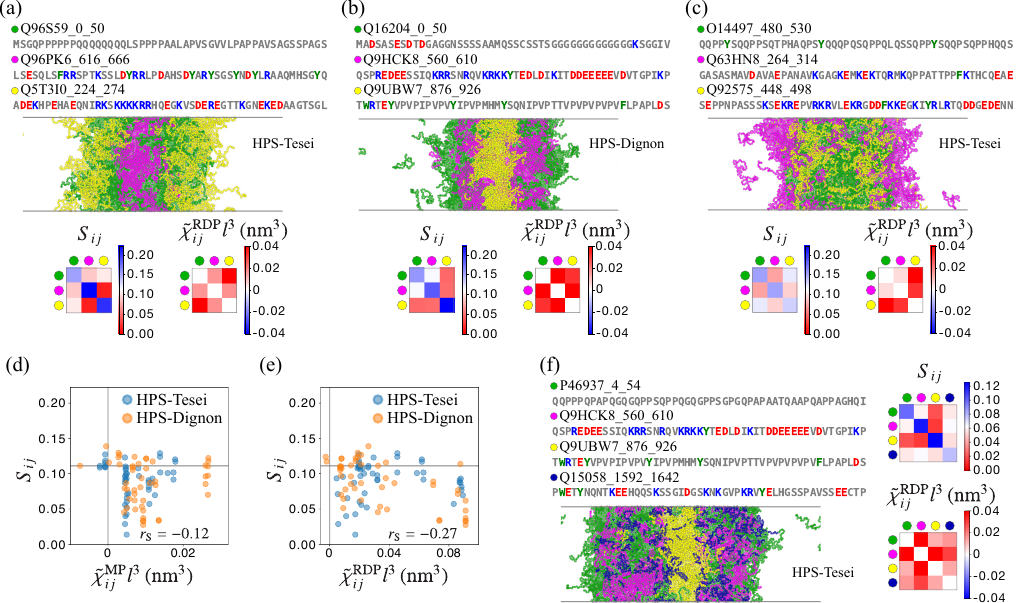}
    \caption{Prediction of demixing for three-component IDR polymers.
    (a-c) The upper panels show snapshots of the simulated three-component IDR sequences.
    The sequences with their names are shown above each configuration.
    The left and middle cases show the demixing of all the components, and the right case is where one of the components (magenta) is well separated from the other two. The extent of separation is quantified in the bottom left panels of the demixing matrix elements, $S_{ij}$.
    Compared to the random configuration where $S_{ij} = 1 / 9$ (white), $S_{ij} < 1 / 9$ (red) suggests that sequences $i$ and $j$ are spatially distant from each other (i.e., demixed if $i \neq j$), and $S_{ij} > 1 / 9$ (blue) suggests that sequences $i$ and $j$ tend to be neighbors.
    The bottom right panels represent the effective inter-component interaction parameter, $\tilde{\chi}_{ij}^\mathrm{RDP}$ (at $T = 300 \ \mathrm{K}$).
    Sequences with positively large $\tilde{\chi}_{ij}^\mathrm{RDP}$ (for $i \neq j$), shown with red color, are expected to undergo demixing and were used for simulations.
    (d, e) The demixing score, $S_\mathrm{demix}$, against the effective inter-component interaction parameter at 300 K obtained by the (d) monomer pair or (e) rescaled dimer pair approximation for 32 sequence triplets in total using the HPS-Tesei and HPS-Dignon models.
    The gray horizontal and vertical lines represent $S_{ij} = 1 / 9$ and $\tilde{\chi}_{ij}^\mathrm{MP} = 0$ (or $\tilde{\chi}_{ij}^\mathrm{RDP} = 0$), respectively.
    Spearman's rank correlation coefficient $r_S$ is shown in each panel.
    (f) Snapshot, $S_{ij}$, and $\tilde{\chi}_{ij}^\mathrm{RDP}$ (at $T = 300 \ \mathrm{K}$) for four-component IDR polymers.}
    \label{Fig:triplet}
\end{figure*}

\section{Effective interaction parameter across distinct sequences}
\label{Sec:predict_demixing}

As the approximated \secVC [Eq.~\eqref{Eq:excluded_volume_dimer_approx}] is useful in predicting the Boyle and critical temperatures, we expect that the \secVC can also account for two-body interactions between distinct IDR polymers.
To capture the interactions between different polymer types ($i$ and $j$), we introduce a matrix that summarizes the pairwise interactions between their dimer units:
\begin{equation}
    B^\mathrm{RDP}_{ij} (T) := \sum_{n = 1}^{N - 1} \sum_{m = 1}^{N - 1} v^\mathrm{RDP}_{d^i_{n, n + 1} d^j_{m, m + 1}} (T).
    \label{Eq:excluded_volume_matrix_dimer_approx}
\end{equation}
Here, $v^\mathrm{RDP}_{d^i_{n, n + 1} d^j_{m, m + 1}}$ is the \secVC for a dimer pair:
\begin{equation}
    v^\mathrm{RDP}_{d^i_{n, n + 1} d^j_{m, m + 1}} (T) := 2 \pi \int _0 ^\infty dr \, r^2 \left[1 - e^{- U^\mathrm{RDP}_{d^i_{n, n + 1} d^j_{m, m + 1}} (r) / T} \right],
    \label{Eq:excluded_volume_dimer_rescaled_muticomponent}
\end{equation}
with $U^\mathrm{RDP}_{d^i_{n, n + 1} d^j_{m, m + 1}}$ being the rescaled dimer-dimer interaction,
\begin{equation}
    U^\mathrm{RDP}_{d^i_{n, n + 1} d^j_{m, m + 1}} := \alpha (U_{a^i_n a^j_m} + U_{a^i_n a^j_{m + 1}} + U_{a^i_{n + 1} a^j_m} + U_{a^i_{n + 1} a^j_{m + 1}}),
    \label{Eq:dimer_interaction_rescaled_multicomponent}
\end{equation}
where $a^i_n$ is the amino acid type of the $n$th monomer in the polymer labeled by $i$.

To understand if different polymer types will separate into distinct phases, we introduce the effective interaction parameter:
\begin{equation}
    \tilde{\chi}^\mathrm{RDP}_{ij} (T) := \frac{B^\mathrm{RDP}_{ij} (T)}{N_i N_j l^3} - \frac{1}{2} \left[ \frac{B^\mathrm{RDP}_{ii} (T)}{{N_i}^2 l^3} + \frac{B^\mathrm{RDP}_{jj} (T)}{{N_j}^2 l^3} \right],
    \label{Eq:B_nondiagonal}
\end{equation}
which quantifies the inter-component interaction strength between components $i$ and $j$ relative to the intra-component interaction strength.
Here, we assumed the correspondence between the interaction parameter and the virial coefficient, $\chi_{ij} = B_{ij} / (N_i N_j l^3) - 1/2$, as used in the prediction of $T_c$ (see Sec.~\ref{Subsec:critical_temperature}), and defined the effective interaction parameter as $\tilde{\chi}_{ij} := \chi_{ij} - (\chi_{ii} + \chi_{jj})/2$.
If the effective interaction parameter $\tilde{\chi}^\mathrm{RDP}_{ij}$ between two sequences is largely negative, they are likely to attract each other. Conversely, a large positive value indicates repulsion.

In Fig.~\ref{Fig:demixing}(a), we plot the distribution of $\tilde{\chi}^\mathrm{RDP}_{ij}$ ($= \tilde{\chi}_{ji}^\mathrm{RDP}$) for some pairs of distinct IDR sequences ($i \neq j$).
For the temperature used in the evaluation of the interaction parameter, we chose $T = \min \{ T_{c, i}, T_{c, j} \}$, where $T_{c, i}$ is the critical temperature obtained by single-component MD simulations for sequence $i$.
We find that most of the effective interactions are negative and positive interactions are rare, suggesting that randomly selected pairs of IDRs tend to have a high affinity with each other.
As we will discuss in Sec.~\ref{SubSec:predict_demixing_eigenspectrum}, the typicality of high affinity across IDR sequences comes from the strong interaction yielded by the charge difference.
We observe that the temperature dependence of $\tilde{\chi}^\mathrm{RDP}_{ij} (T)$ is low (Fig.~\ref{Fig:Btilde_Tdep} in Appendix), making it a robust indicator of repulsiveness between the two polymers.

Another method called RPA also allows the calculation of interactions between distinct polymers in a sequence-dependent manner~\cite{lin2017charge}.
In RPA, where only Coulomb interactions are typically considered, the interaction strength between two different polymers depends only on their individual self-interactions, irrespective of the choice of the residue-level interaction potential.
This results in the $\chi$ parameter becoming a rank one matrix~\cite{jacobs2023theory}, satisfying the geometric mean rule, $\chi_{ij}^2 \sim \chi_{ii} \chi_{jj} $~\cite{lin2017charge}.
It then follows that $\tilde{\chi}_{ij}:= \chi_{ij} - (\chi_{ii} + \chi_{jj})/2$ is close to zero for all pairs of sequences, unless one of the self-interaction parameters is significantly larger than the other (e.g., $\chi_{ii} \gg \chi_{jj}$).
In the case of $\chi^\mathrm{RDP}_{ij}$, on the other hand, the geometric mean rule is clearly violated even without this imbalance in the self-interactions, as shown in Fig.~\ref{Fig:compare_rpa} in Appendix, indicating that the predicted interaction strength is distinct from the RPA results.
As we will see in Section~\ref{SubSec:predict_demixing_eigenspectrum}, the $\chi$ parameters calculated for the monomer pair and dimer pair approximations are also low rank, except that there are at least several non-zero eigenvalues, which is why the effective interaction parameter can deviate from zero.

\subsection{Predicting demixing and hypermixing in two-component simulations}
\label{SubSec:predict_demixing_two_comp}

To see how $\tilde{\chi}^\mathrm{RDP}_{ij}$ relates to the behavior of condensates, we conducted MD simulations of mixtures containing two types of sequences.
The simulation was conducted similarly to the critical temperature estimation but with $200$ molecules each and for $500\text{ ns}$.
For the temperature, we chose $ \min \{ T_{c, i}, T_{c, j} \}$ rounded down to the nearest 10 K, where $T_{c, i}$ is the critical temperature obtained by single-component MD simulations for a sequence $i$.

In Fig.~\ref{Fig:demixing}(b), we show examples of snapshots for equilibrated two-component polymers (green and magenta).
For a negatively large $\tilde{\chi}_{ij}^\mathrm{RDP} $ [$= -0.56 \text{ nm}^3/l^3$, bottom left of Fig.~\ref{Fig:demixing}(b)], we find that polymers show condensation with a highly uniform distribution of the two components, which we call a hypermixed condensate, suggesting that the inter-component affinity exceeds the intra-component affinity.
In contrast, for a positively large $\tilde{\chi}_{ij}^\mathrm{RDP} $ [$= 0.060 \text{ nm}^3/l^3$, top of Fig.~\ref{Fig:demixing}(b)], the two components are demixed; the two species form separate condensates that do not mix.
For an intermediate $\tilde{\chi}_{ij}^\mathrm{RDP} $ [$= -0.017 \text{ nm}^3/l^3$, bottom right of Fig.~\ref{Fig:demixing}(b)], the two components seem to be mixed randomly within a single condensate.

To quantify the degree of separation between polymer types, we developed the demixing score $S_\mathrm{demix}$, which measures how often neighboring polymer chains belong to the same type (see Appendix \ref{App:multicomponent}).
A demixing score $S_\mathrm{demix}$ of $0.5$ indicates a well-mixed state where the two polymer types are randomly distributed. A lower score suggests they prefer to be near each other (hypermixing), while a higher score indicates separation (demixing). This is confirmed for each configuration in Fig.~\ref{Fig:demixing}(b).

To examine the general predictability of demixing by $\tilde{\chi}_{ij}^\mathrm{RDP}$, we further performed multiple simulations and calculated  $S_\mathrm{demix}$ for two-component polymer systems for the HPS-Tesei~\cite{tesei2021accurate} and HPS-Dignon~\cite{dignon2018relation} models.
As shown in Fig.~\ref{Fig:demixing}(c), the distributions of $S_\mathrm{demix}$ for negatively large $\tilde{\chi}_{ij}^\mathrm{RDP} $ ($< -0.04  \text{ nm}^3 /l^3$), near-zero $\tilde{\chi}_{ij}^\mathrm{RDP} $ ($\in [-0.04, 0.04] \text{ nm}^3 /l^3$), and positively large $\tilde{\chi}_{ij}^\mathrm{RDP} $ ($> 0.04 \text{ nm}^3 /l^3$) pairs are distinct, confirming that larger $\tilde{\chi}_{ij}^\mathrm{RDP}$ indicates larger $S_\mathrm{demix}$, i.e., a higher tendency toward demixing.
Spearman's rank correlation coefficient between $S_\mathrm{demix}$ and $\tilde{\chi}_{ij}^\mathrm{RDP}$ is $r_S = 0.67$.
Although the correlation is lower than $\tilde{\chi}_{ij}^\mathrm{RDP}$, we find that $\tilde{\chi}_{ij}^{\mathrm{MP}}$, the effective interaction parameter calculated from the monomer pair approximation:
\begin{equation}
    \tilde{\chi}^\mathrm{MP}_{ij} (T) := \frac{B^\mathrm{MP}_{ij} (T)}{N_i N_j l^3} - \frac{1}{2} \left[ \frac{B^\mathrm{MP}_{ii} (T)}{{N_i}^2 l^3} + \frac{B^\mathrm{MP}_{jj} (T)}{{N_j}^2 l^3} \right],
    \label{Eq:B_nondiagonal_monomer}
\end{equation}
also positively correlates with $S_\mathrm{demix}$ ($r_S = 0.52$, see Fig.~\ref{Fig:demixing_supplement} in Appendix).
As we will show in Sec.~\ref{SubSec:predict_demixing_antagonistic}, $\tilde{\chi}_{ij}^{\mathrm{MP}}$ is useful in practice when designing demixing sequences.

\begin{figure*}[t]
    \centering
    \includegraphics[scale=1]{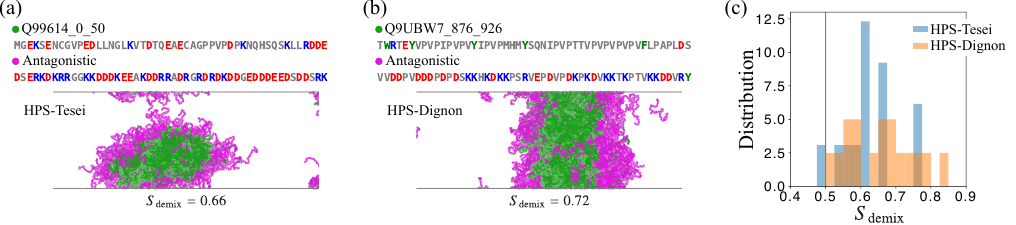}
    \caption{Demixing of IDR polymers and the generated antagonistic polymers.
    (a, b) Snapshots of the simulated IDR (green) and antagonistic (magenta) sequences shown with the sequence and its name.
    (c) The distribution of the demixing score $S_\mathrm{demix}$ for the pairs of IDR and antagonistic sequences with positively large $\tilde{\chi}_{ij}^\mathrm{MP}$.
    We tested 13 sequence pairs (with mean $\tilde{\chi}_{ij}^\mathrm{MP}l^3 = 0.022 \text{ nm}^3$) using the HPS-Tesei model and 16 sequence pairs (with mean $\tilde{\chi}_{ij}^\mathrm{MP}l^3 = 0.024 \text{ nm}^3$) using the HPS-Dignon model.}
    \label{Fig:demixing_antagonistic}
\end{figure*}

We note that other quantities have been introduced to measure the extent of demixing, such as $S_\mathrm{proj}:= |\rho_{i,\mathrm{center}}-\rho_{j,\mathrm{center}}|$ in Ref.~\cite{chew2023thermodynamic}. Here, $\rho_{i,\mathrm{center}}$ is the number density of component $i$ at the center of the condensate, calculated using the number density projected onto the $z$-axis. 
As expected, the two scores, $S_\mathrm{demix}$ and $S_\mathrm{proj}$ correlate significantly when calculated for the configurations obtained in the two-component simulations.
We here nevertheless continue using $S_\mathrm{demix}$ over $S_\mathrm{proj}$ since it can discriminate hypermixing states from randomly mixed states (see the region corresponding to $S_\mathrm{demix} \simeq 0.5$ and below in Fig.~\ref{Fig:rev_demix} in Appendix). 
There is also the advantage that $S_\mathrm{demix}$ does not rely on the projection of the density to an axis; it can probe the segregation of components even when the condensates have not reached a one-dimensional profile, due for example to the aggregate state~\cite{rana2021phase} (see bottom right examples in Fig.~\ref{Fig:rev_demix}).
The demixing score $S_\mathrm{demix}$ can also be generalized to multi-component cases, as we will see later.

Theoretical and experimental analysis regarding the two-component Flory-Huggins theory has pointed out that the angle of the tie line can indicate whether two components will mix or separate~\cite{sear2003instabilities,jacobs2013predicting,jacobs2017phase,qian2022tie}.
This scheme predicts that the extent of demixing will depend on the raw $B_{ij}$, or specifically the angle of the tie line which can be approximated by the angle $\bar{\theta}_{ij} \in [-\pi / 2, \pi / 2)$ of the eigenvector (measured from the first axis) for the smaller eigenvalue of the $2 \times 2$ matrix,
\begin{equation}
    \begin{pmatrix}
        B_{ii}(T) & B_{ij}(T) \\
        B_{ji}(T) & B_{jj}(T)
    \end{pmatrix},
\end{equation}
when assuming equal density.
To remove the arbitrariness in choosing the first axis, we define an angle $\theta_{ij} := -\mathrm{sgn}(\bar{\theta}_{ij}) \min \{ |\bar{\theta}_{ij}|, \pi/2 - |\bar{\theta}_{ij}| \}$ ($\in [-\pi / 4, \pi / 4]$), which is expected to be positive in the case of demixing and negative in the case of mixing.
We tested how these quantities correlate with the demixing score using the estimated $B^\mathrm{RDP}(T)$ and $B^\mathrm{MP}(T)$ at $T = \min \{ T_{c, i}, T_{c, j} \}$, and found that the effective interaction parameters [Eqs.~\eqref{Eq:B_nondiagonal} and \eqref{Eq:B_nondiagonal_monomer}] perform better in terms of predicting demixing [Figs.~\ref{Fig:demixing_supplement}(e) and (f) in Appendix].

\subsection{Demixing more than two components}
\label{SubSec:predict_demixing_M}

We then explored whether our method could predict the separation of mixtures containing more than two types of polymer ($M>2$), where each pair of components $i$ and $j$ has a positive value for $\tilde{\chi}_{ij}$ (see Appendix~\ref{App:multicomponent} for the method to select the sequences).

For mixtures with three polymer types ($M=3$), we found that some of the predicted set of sequences indeed undergo demixing, as demonstrated in Figs.~\ref{Fig:triplet}(a) and (b) for the HPS-Tesei and HPS-Dignon models, respectively.
The extent of demixing is quantified by the demixing matrix $\bm{S}$ whose components represent the fraction of edges within the $k$-nearest neighbor graph, which is normalized as $\sum_{1\leq i,j \leq M} S_{ij}=1$.
The demixing matrix is the generalization of the demixing score as it satisfies $S_\mathrm{demix}= \mathrm{Tr} \, \bm{S}$ for $M=2$, and should become $S_{ij}=1/M^2$ when the configuration of polymers is random.
As seen in the examples [Figs.~\ref{Fig:triplet}(a-c)], the large positive values of $\tilde{\chi}_{ij}^\mathrm{RDP}$ for each set of sequences lead to demixing matrix components that are lower than random, $1/M^2=1/9$, although mixed pairs of polymers ($>1/9$) can appear in some cases [Fig.~\ref{Fig:triplet}(c)].
Testing with 32 triplets for both the HPS-Tesei and HPS-Dignon models, we find the trend that $\tilde{\chi}_{ij}^\mathrm{RDP}$ is negatively correlated with $S_\mathrm{demix}$, more significantly compared with $\tilde{\chi}_{ij}^\mathrm{MP}$ [Figs.~\ref{Fig:triplet}(d) and (e)].

We further tested if we can achieve demixing of $M=4$ sequences using similar criteria for selecting sequences.
Unfortunately, we could not find a demixing quadruplet within the set of IDRs that we tested; in all the 16 cases that we have tried, we obtained three phases with one including two species of polymers [Fig.~\ref{Fig:triplet}(f)].
This can be due to the limitation of our IDR library, as we could not find quadruplets of which all the pairs within it have large positive values of $\tilde{\chi}_{ij}^\mathrm{RDP}$, but can also be due to the theoretical bound in the number of demixing components, as we discuss in Section \ref{Sec:max_demix}.

\begin{figure*}[t]
    \centering
    \includegraphics[scale=1]{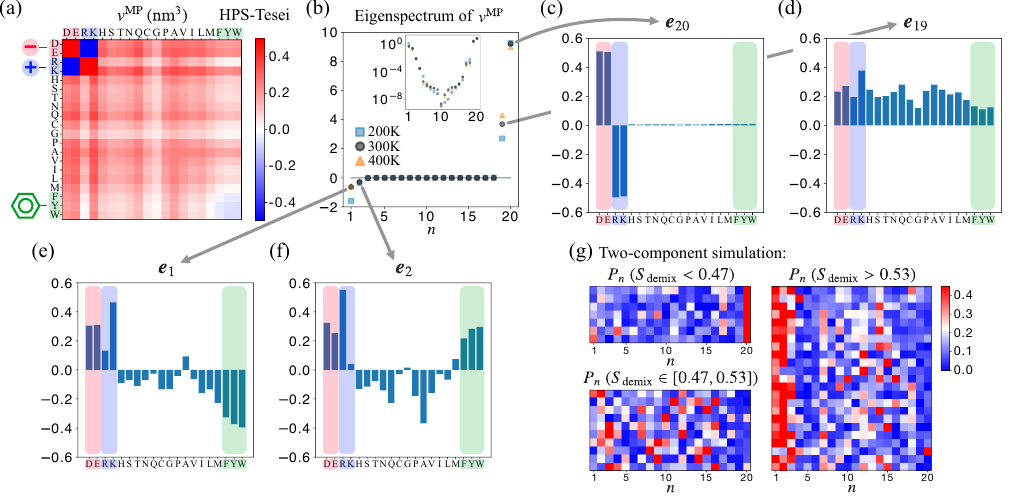}
    \caption{Molecular insights for the mechanism of demixing and hypermixing. (a) \secVC matrix calculated at $300\text{ K}$ in the monomer pair approximation using the HPS-Tesei parameters. (b) Eigenspectrum of the \secVC matrix. Inset shows the absolute values of the eigenvalues with logscale. (c) Eigenvector with the largest eigenvalue. (d) Eigenvector with the second largest eigenvalue. (e) Eigenvector with the most negative eigenvalue. (f) Eigenvector with the second most negative eigenvalue. For all the plots of eigenvectors, the component for amino acid D is taken as positive. (g) Eigendecomposition of the amino acid fraction difference vector $\bm{n}_i - \bm{n}_j$ for the pairs tested in the two-component simulation (Fig.~\ref{Fig:demixing}). The three panels represent the hypermixing ($S_\mathrm{demix}<0.47$), random ($S_\mathrm{demix} \in [0.47,0.53]$), and demixing ($S_\mathrm{demix} > 0.53$)  pairs.}
    \label{Fig:exvol_matrix}
\end{figure*}

\subsection{Generation of antagonistic sequence for demixing}
\label{SubSec:predict_demixing_antagonistic}

We can use our approximation scheme to design new polymer sequences that will repel and separate from a specific given sequence.
The results in Secs.~\ref{SubSec:predict_demixing_two_comp} and \ref{SubSec:predict_demixing_M} suggest that $\tilde{\chi}_{ij}^\mathrm{MP}$ and $\tilde{\chi}_{ij}^\mathrm{RDP}$ can work as indicators of the tendency toward demixing.
Here, we use $\tilde{\chi}_{ij}^\mathrm{MP}$ for simplicity as well as for interpretability, as it can be calculated solely from the amino acid compositions.

For a given IDR sequence $i$, the aim is to generate a second sequence $j$ that demixes with it. 
To observe demixing at the temperature regime comparable to the condensing regime for sequence $i$, we maximize $\tilde{\chi}_{ij}^\mathrm{MP}(T)$ calculated at $T=T_{c,i}$, the critical temperature for sequence $i$, while constraining $\chi_{ii}^\mathrm{MP} (T_{c,i}) = \chi_{jj}^\mathrm{MP} (T_{c,i})$.
Without the constraint, we would typically obtain sequence $j$ with $\tilde{\chi}_{ij}^\mathrm{MP}(T_{c,i})$ and $\chi_{jj}^\mathrm{MP}(T_{c,i})$ being both large, in which case sequence $j$ will not undergo phase separation at $T_{c,i}$.
We also set $N_\mathrm{replace}$, the total number of amino acid compositional differences between sequences $i$ and $j$.

We can represent each sequence as a vector, $\bm{n}_i$, where each element $n_{i, a}$ corresponds to the proportion of a specific amino acid type $a$ contained in the sequence $i$, and satisfying $\sum_a n_{i, a} = 1$. 
Using this representation, the effective interaction parameter can be expressed as
\begin{equation}
\tilde{\chi}_{ij}^\mathrm{MP} = - \frac{1}{2 l^3} (\bm{n}_i - \bm{n}_j)^\mathrm{T} v^\mathrm{MP} (\bm{n}_i - \bm{n}_j).
\label{Eq:B_nondiagonal_vec}
\end{equation}
Here, $v^\mathrm{MP}$ is a $20 \times 20$ matrix with the elements defined by Eq.~\eqref{Eq:excluded_volume_monomer}.
The formula~\eqref{Eq:B_nondiagonal_vec} indicates that $\tilde{\chi}_{ij}^\mathrm{MP}$ only depends on the difference in the amino acid fraction between the sequences.
The three constraints we impose are
\begin{equation}
    N_i=N_j=N,
    \label{Eq:maximize_B12_constraint1}
\end{equation}
\begin{equation}
    {\bm{n}_i}^\mathrm{T} v^\mathrm{MP} \bm{n}_i = {\bm{n}_j}^\mathrm{T} v^\mathrm{MP} \bm{n}_j,
\end{equation}
and
\begin{equation}
    \sum_a |N_{i, a} - N_{j, a}| = 2 N_\mathrm{replace}.
    \label{Eq:maximize_B12_constraint3}
\end{equation}

To maximize $\tilde{\chi}_{ij}^\mathrm{MP} (T_{c,i})$ with respect to $\bm{n}_j$ under the constraints~\eqref{Eq:maximize_B12_constraint1}-\eqref{Eq:maximize_B12_constraint3}, we employed a Python package for optimization (scipy.optimize.differential\_evolution~\cite{virtanen2020scipy}) using the default parameters, with $N = 50$ and $N_\mathrm{replace} = 25$.
We rounded decimals of the obtained elements of $N_j \bm{n}_j$ and added or removed randomly chosen amino acids so that the length of the sequence becomes 50.
We generated sequence $j$ by randomly shuffling the order of amino acids from this sequence.

In Figs.~\ref{Fig:demixing_antagonistic}(a) and (b), we show simulation results of the obtained sequences that indeed show demixing with the given sequences.
For 13 (16) pairs of given and generated sequences with positively large $\tilde{\chi}_{ij}^\mathrm{MP}$ that are tested using the HPS-Tesei (HPS-Dignon) model, we show the distribution of the demixing score $S_\mathrm{demix}$ in Fig.~\ref{Fig:demixing_antagonistic}(c).
The distribution suggests that the proposed approach is useful in generating the antagonistic sequence that will demix with a given sequence;  most of them have $S_\mathrm{demix} > 0.5$ and the majority achieves $S_\mathrm{demix} > 0.6$.

\begin{figure*}[t]
    \centering
    \includegraphics[scale=1]{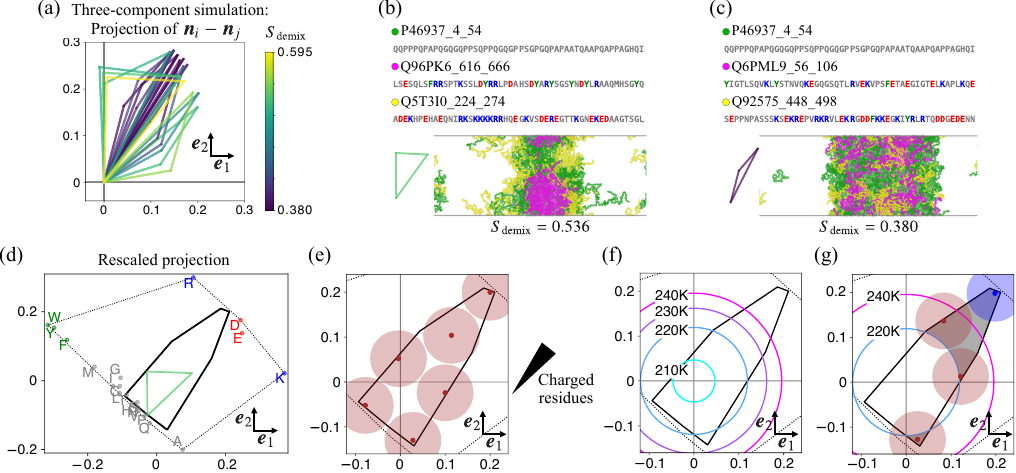}
    \caption{Region spanned by the sequence vectors in the demixing plane. (a) Projection of the difference vectors of the pairs of fractions in the tested triplets in the two-dimensional space spanned by $\bm{e}_1$ and $\bm{e}_2$ (demixing plane) for the HPS-Tesei model (see Fig.~\ref{Fig:exvol_matrix}). We translated the position of each triangle for comparison. (b, c) Examples of (b) high demixing score and (c) relatively low demixing score with the corresponding triangles. (d) Projection of the single amino acid vectors onto the demixing plane with the axes rescaled by the eigenvalues. The region surrounded by the dotted line is the convex hull calculated from the 20 single amino acid vectors, which corresponds to the region explorable by all the sequences. The region defined by the bold line is obtained by enforcing $\tilde{n}_{i,19}=0.24 \sqrt{\lambda_{19}} = 0.46 \text{ nm}^{3 / 2} \, (=\text{const.})$ and  $\tilde{n}_{i,20}=0$. The value for $\tilde{n}_{i,19}$ was selected to roughly maximize the area of the space after the restriction. The triangle corresponding to the sequence triplet in (b) is shown to confirm that it fits within this region. (e) The possible combination of sequences that can be packed within the restricted space in the demixing plane when setting the minimum distance between the points as $\sqrt{2 \tilde{\chi}^\mathrm{MP}_\mathrm{min}l^3}$ with $\tilde{\chi}^\mathrm{MP}_\mathrm{min}l^3=0.008 \text{ nm}^3$. This number $M=6$ seems to be the maximum for this condition. (f) The same restricted space in the demixing plane with circles representing $T_c \approx \text{const.}$, as suggested by the temperature written near each circle. Here, the relation between $T_c$ and $\chi_{ii}^\mathrm{MP}$ is estimated by linear fitting (see Appendix~\ref{App:estimate_critical_temp_from_chimp}). (g) If enforcing the condition that all $M$ sequences should be sitting inside the restricted space, separated by at least $\sqrt{0.016} \text{ nm}^{3 / 2}$ from each other, and should be sitting between two circles that correspond to $T_c \approx \text{220 K}$ and $T_c \approx \text{240 K}$, then $M=3$ is the maximum; without the constraint on the upper limit of $T_c$, $M=4$ is the maximum (e.g., a sequence corresponding to the blue point is allowed).}
    \label{Fig:triplet_projection}
\end{figure*}

\subsection{Rules of demixing and hypermixing deduced from the eigenspectrum of \secVC matrix}
\label{SubSec:predict_demixing_eigenspectrum}

From the formula of $\tilde{\chi}_{ij}^\mathrm{MP}$ in Eq.~\eqref{Eq:B_nondiagonal_vec}, we can understand what drives two sequences to mix or separate by analyzing $v^\mathrm{MP}$ and the differences in amino acid composition between two sequences, $\bm{n}_i - \bm{n}_j$.
In Figs.~\ref{Fig:exvol_matrix}(a) and (b), we show $v^\mathrm{MP}$ calculated at $T=300 \ \mathrm{K}$ and its eigenspectrum $\{ \lambda_n \}_{n = 1}^{20}$ for the HPS-Tesei model, respectively.
As expected, the charged residues dominate the interactions, which is seen as the largest eigenvalue $\lambda_{20}$ and its corresponding eigenvector $\bm{e}_{20}$ [Fig.~\ref{Fig:exvol_matrix}(c)], where $\bm{e}_n$ is the normalized eigenvector for the $n$th smallest eigenvalue $\lambda_n$.
This explains why many pairs tended to have low $\tilde{\chi}_{ij}^\mathrm{MP}$ (Fig.~\ref{Fig:demixing_dist_supplement} in Appendix) as well as $\tilde{\chi}_{ij}^\mathrm{RDP}$ [Fig.~\ref{Fig:demixing}(a)] and therefore tended to hypermix as shown in the bottom left panel of Fig.~\ref{Fig:demixing}(b); difference in charge is itself already a strong driver of co-condensation.

On the other hand, demixing is rare since $\bm{n}_i - \bm{n}_j$ must align closely with the few eigenvectors that correspond to negative eigenvalues of $v^\mathrm{MP}$.
For the HPS-Tesei model, $\bm{e}_1$ [Fig.~\ref{Fig:exvol_matrix}(e)] indicates that demixing from a condensing sequence $i$ can be achieved by making sequence $j$ by exchanging charged residues (marked by the red and blue shadows) with aromatic residues (marked by the green shadow) from sequence $i$, or vice versa.
This is a plausible strategy given that charged residues and aromatic residues are indeed both known to be abundant in proteins forming condensates, and these condensates seem to attract distinct components inside cells~\cite{lyons2023functional}.

The second eigenvector $\bm{e}_2$ [Fig.~\ref{Fig:exvol_matrix}(f)] indicates that exchanging charged {\it and} aromatic residues with other weaker interacting residues (A, Q, P, etc.) can also lead to demixing.
To examine if the demixing and hypermixing pairs found in simulations respect these eigenvectors, we compared $\bm{n}_i - \bm{n}_j$ for the $62$ sequence pairs used in plotting Fig.~\ref{Fig:demixing}(c) for the HPS-Tesei model.
In Fig.~\ref{Fig:exvol_matrix}(g), we plot $P_n$, calculated by
\begin{equation}
    P_n := |\bm{e}_n \cdot (\bm{n}_i - \bm{n}_j)| / || \bm{n}_i - \bm{n}_j ||,
\end{equation}
as a heatmap for hypermixed ($S_\mathrm{demix} < 0.47$), randomly mixed ($S_\mathrm{demix} \in [0.47, 0.53]$), and demixed ($S_\mathrm{demix} > 0.53$) sequence pairs (see Fig.~\ref{Fig:projection_all} for $P_n$ and $S_\mathrm{demix}$ for all the sequence pairs).
We find that most of the demixing pairs [$S_\mathrm{demix} > 0.53$, right panel of Fig.~\ref{Fig:exvol_matrix}(g)] have the fraction difference vector that has large components of not only $\bm{e}_1$ but also $\bm{e}_2$, highlighting the fact that both directions, including their mixtures, can be practical ways in achieving demixing.

The hypermixing pairs ($S_\mathrm{demix}<0.47$) have large $\bm{e}_{20}$ [Fig.~\ref{Fig:exvol_matrix}(c)] components as expected, whereas neutrally interacting pairs ($S_\mathrm{demix} \in [0.47,0.53]$) have nonzero components scattered within the eigenvectors corresponding to close-to-zero eigenvalues [left panels of Fig.~\ref{Fig:exvol_matrix}(g)].
Here, none of the fraction difference vectors have a large $\bm{e}_{19}$ [Fig.~\ref{Fig:exvol_matrix}(d)] component since $\bm{e}_{19}$ is close to uniform, which means that the difference vector is approximately orthogonal to it by definition [i.e., $\sum_a (n_{i, a} - n_{j, a})=0$].
The effective interaction parameter has the useful property that it does not depend strongly on the temperature (Fig.~\ref{Fig:Btilde_Tdep} in Appendix); this can be explained by the weak temperature dependence of the eigenspectrum as shown with different marks in Fig.~\ref{Fig:exvol_matrix}(b) except for the 19th eigenvalue, which is irrelevant as explained.

We conducted the same eigenspectrum analysis on $v^\mathrm{MP}$ from the HPS-Dignon parameter set (Fig.~\ref{Fig:exvol_matrix_dignon} in Appendix) as well as the Mpipi model (Fig.~\ref{Fig:exvol_matrix_mpipi} in Appendix).
Apart from the difference in the charge of histidine ($0$ in HPS-Tesei, $+0.5$ in HPS-Dignon, and $+0.375$ in Mpipi), the situation is similar to the HPS-Tesei model in the positive (hypermixing) eigenvectors. For the negative (demixing) eigenvectors, the Mpipi model has $\bm{e}_1$ and $\bm{e}_2$ that treat the combination of aromatic residues and charged residues in a distinct manner, highlighting the difference between arginine (R) and lysine (K) in terms of the interaction with the aromatic residues (F, Y, and W). 
Nevertheless, the prominent feature that there are at most only two vectors that span the demixing space is shared across the three models. Previous studies have also highlighted the limited number of ways that amino acids can interact with each other~\cite{li1997nature,graf2022thermodynamic}

\subsection{Maximum number of demixing sequences}
\label{Sec:max_demix}
Seeing that the number of demixing vectors is limited (in fact, only two), we wondered how demixing of more than two components can be achieved. 
In the three-component simulations (Fig.~\ref{Fig:triplet}), the chosen pairs had large values of $\bm{e}_1$ and $\bm{e}_2$ components in the difference vectors $\{ \bm{n}_i - \bm{n}_j \}$ [Fig.~\ref{Fig:triplet_projection}(a)].
For the mixtures where the three polymer types separated well, their corresponding points projected on the space spanned by $\bm{e}_1$ and $\bm{e}_2$ formed a larger triangle [Figs.~\ref{Fig:triplet_projection}(a-c)].
This indicates that within this two-dimensional space, the difference vectors can utilize the combination of distinct directions to achieve demixing of more than two components.

For multiple polymer types to separate, the effective interaction parameter $\tilde{\chi}^\mathrm{MP}_{ij}$ between each pair must be sufficiently large to drive them apart.
To see if there is a theoretical maximum in the number of components that are demixed, here we seek a geometrical representation of the problem.

By introducing $\tilde{n}_{i,n} := \sqrt{| \lambda_n|} \bm{e}_n \cdot \bm{n}_i$, the effective interaction parameter can be approximated by
\begin{equation}
\tilde{\chi}^\mathrm{MP}_{ij} \simeq \frac{1}{2l^3} \sum_{n=1,2}(\tilde{n}_{i,n} - \tilde{n}_{j,n})^2 -\frac{1}{2l^3} \sum_{n=19,20}(\tilde{n}_{i,n} - \tilde{n}_{j,n})^2,
\label{Eq:B_nondiagonal_vec_approx}
\end{equation}
since $\lambda_1$ and $\lambda_2$ are negative, $\lambda_{19}$ and $\lambda_{20}$ are positive, and $\lambda_n \simeq 0$ for all other $n$ [Fig.~\ref{Fig:exvol_matrix}(b)].
Seeing this representation, we observe that $\tilde{\chi}^\mathrm{MP}_{ij}$ can become large by setting the Euclidian distance between the sequences large in the demixing space (spanned by $\bm{e}_1$ and $\bm{e}_2$) while minimizing the Euclidian distance in the hypermixing space (spanned by $\bm{e}_{19}$ and $\bm{e}_{20}$).

Any fraction vector can be written as $\bm{n}_i = \sum_a n_{i,a} \bm{e}_a$, where $\bm{e}_a$ is the single amino acid vector with the component at amino acid $a$ being one and all other components zero and $n_{i,a} \geq 0$.
The projection $\tilde{n}_{i,n}$ is written as $\tilde{n}_{i,n} = \sum_a n_{i,a}  \tilde{e}_{a,n}$, where $\tilde{e}_{a,n}:= \sqrt{|\lambda_n|} \bm{e}_n \cdot \bm{e}_a$ is the projection of the single amino acid vector. The point $(\tilde{n}_{i,1},\tilde{n}_{i,2})$ is then restricted within the convex hull that is formed by the 20 points of projections $\{ (\tilde{e}_{a,1}, \tilde{e}_{a,2}) \}_a$ in the demixing space [Fig.~\ref{Fig:triplet_projection}(d), dotted line].
We find that the top right region in this convex hull is spanned by the charged residues, the left region is dominated by the aromatic residues, and the bottom region is dominated by the others including proline and glutamine.

A further constraint is set by restricting $\tilde{n}_{i,19}$ and $\tilde{n}_{i,20}$ to be all the same across sequences to make the second term in Eq.~\eqref{Eq:B_nondiagonal_vec_approx} to be minimal.
The possible region within the demixing plane that the sequences can take is then calculated by first obtaining the convex hull formed by 20 points $\{ (\tilde{e}_{a,1}, \tilde{e}_{a,2}, \tilde{e}_{a,19}, \tilde{e}_{a,20}) \}_a$ in the four-dimensional space and obtaining the cross-section with $\tilde{n}_{i,19}=\mathrm{const.}$ and $\tilde{n}_{i,20}=\mathrm{const.}$ (see Appendix~\ref{App:projection} for details).
The resulting region is shown as the bold line region in Fig.~\ref{Fig:triplet_projection}(d), where we took $\tilde{n}_{i,19}=0.24 \sqrt{\lambda_{19}} = 0.46 \text{ nm}^{3 / 2}$ and $\tilde{n}_{i,20}=0$ [i.e., restriction to almost charge neutral sequences as seen from $\bm{e}_{20}$ in Fig.~\ref{Fig:exvol_matrix}(c)].

Finding multiple polymer types that demix with each other is then equivalent to placing points within a confined area, ensuring that they are sufficiently far apart from each other.
If the minimum value required for $\tilde{\chi}^\mathrm{MP}_{ij}$ to demix is $\tilde{\chi}^\mathrm{MP}_{\text{min}}$, we need to place points keeping the distance of at least $\sqrt{2 \tilde{\chi}^\mathrm{MP}_{\text{min}}}$ in the metric of the rescaled projection, according to Eq.~\eqref{Eq:B_nondiagonal_vec_approx}.
In Fig.~\ref{Fig:triplet_projection}(e), we plot non-overlapping circles with diameter $\sqrt{2 \tilde{\chi}^\mathrm{MP}_{\text{min}}}$ within the restricted space corresponding to $\tilde{\chi}^\mathrm{MP}_{\text{min}}l^3=0.008 \text{ nm}^3$.
This figure suggests that $M=6$ is the maximum number of points that can be taken this way.

We can in principle take the points (i.e., sequences) in the upper right area of Fig.~\ref{Fig:triplet_projection}(e) [i.e., sequences with a high percentage of charged amino acids, see also Fig.~\ref{Fig:triplet_projection}(d)].
However, those points will be away from the origin [i.e., $(0,0)$ in the demixing plane], meaning that the diagonal element of $\chi^\mathrm{MP}$,
\begin{equation}
\chi^\mathrm{MP}_{ii} \simeq -\frac{1}{l^3}\sum_{n=1,2}\tilde{n}_{i,n} ^2 + \frac{1}{l^3}\sum_{n=19,20}\tilde{n}_{i,n}^2 - \frac{1}{2},
\label{Eq:B_diagonal_vec_approx}
\end{equation}
would become smaller due to the first term, and therefore have a higher critical temperature $T_{c, i}$.
In Fig.~\ref{Fig:triplet_projection}(f), we show concentric circles in the demixing space that approximately correspond to fixed critical temperatures (see Appendix~\ref{App:estimate_critical_temp_from_chimp} for the derivation).
When restricting the critical temperature for each sequence to be similar to each other, we should place all the points between two circles that represent the upper and lower limits of $T_{c, i}$.
For example, if we assume $220 \text{ K} \leq T_{c,i} \leq 240 \text{ K}$, we find that $M=3$ is the maximum number for demixing under this restriction [three brown dots in Fig.~\ref{Fig:triplet_projection}(g)]; without the upper limit of $T_{c,i}$, $M=4$ is the maximum [additional blue dot in Fig.~\ref{Fig:triplet_projection}(g)].

In summary, we have found that there is a limitation in the number of demixable components when assuming the following points. The first assumption is that the effective interaction parameter calculated from the monomer pair approximation has to be larger than a certain value for the two sequences to demix with each other. This is based on the positive correlation between the effective interaction parameter and the demixing score that we have observed (Fig.~\ref{Fig:demixing} and Fig.~\ref{Fig:demixing_supplement} in Appendix). Seeing that not all triplet simulations that we tested underwent demixing, this assumption is likely a required condition rather than a sufficient condition, meaning that the real restriction should be even stronger. The second assumption is that the components of the hypermixing dimensions ($\bm{e}_{19}$ and $\bm{e}_{20}$) in the difference vector ($\bm{n}_i - \bm{n}_j$) must be suppressed to zero to bring the effective interaction parameter large. This seems unavoidable since a small proportion of these components will bring the effective interaction parameter significantly lower; it is seen that the $\bm{e}_{19}$ and $\bm{e}_{20}$ components are indeed small for the pairs with high demixing scores [Fig.~\ref{Fig:exvol_matrix}(g)]. The third assumption is to restrict the set of sequences to those that have sufficient self-interactions, which can undergo phase separation at realistic temperatures. Since all sequences will ultimately undergo condensate formation for a low enough temperature, it is unnatural to allow arbitrary levels of self-interactions. Where to set this minimum critical temperature is arbitrary [see the example case depicted in Fig.~\ref{Fig:triplet_projection}(g)]. Nevertheless, we still have the limitation of six sequences as the maximum number of demixable elements, even without the restriction on the critical temperature[Fig.~\ref{Fig:triplet_projection}(e)].

The limitations on the number of demixing components seem to be a general principle, holding true even when we use different models or consider the rescaled dimer pair approximation.
Writing the amino acid dimer fraction of sequence $i$ as $\bm{n}_i^\mathrm{D}$, where the element $n_{i, d}^\mathrm{D}$ is the number of amino acid dimers of type $d$ ($\in \{\mathrm{AA}, \mathrm{AC}, \mathrm{AD}, \cdots, \mathrm{YW}, \mathrm{YY}\}$) contained in sequence $i$, we can express $\tilde{\chi}_{ij}^\mathrm{RDP}$ as
\begin{equation}
\tilde{\chi}_{ij}^\mathrm{RDP} = - \frac{1}{2l^3} (\bm{n}_j^\mathrm{D} - \bm{n}_i^\mathrm{D})^\mathrm{T} v^\mathrm{RDP} (\bm{n}_j^\mathrm{D} - \bm{n}_i^\mathrm{D}).
\label{Eq:B_nondiagonal_vec_dimer}
\end{equation}
Here, $v^\mathrm{RDP}$ is a $400 \times 400$ matrix with the elements constructed using Eq.~\eqref{Eq:excluded_volume_dimer_rescaled_muticomponent}.
The number of dimensions of the demixing space, which is the number of eigenvectors of $v^\mathrm{RDP}$ with negative eigenvalues, is no more than two for the HPS-Tesei, HPS-Dignon, and Mpipi models (Fig.~\ref{Fig:exvol_eigval_dimer} in Appendix), indicating that the restriction in placing points in the demixing space is similar even for the rescaled dimer pair approximation model. 

\section{Discussion and conclusion}
Here we have shown that the prediction of heteropolymer interactions including demixing upon condensation is possible for the coarse-grained models of disordered region sequences. The method we propose allows the estimation of the Boyle temperature, critical temperature, the selection of IDR sequences that will demix or hypermix, as well as the generation of antagonistic (demixing) components for a given sequence.

We have shown how the monomer pair approximation captures the basic properties of the interactions but can be improved quantitatively by the dimer pair approximation.
This indicates that the essential interaction range is slightly beyond a monomer, which is plausible given that the bond length ($l_b = 0.38 \text{ nm}$) is comparable to the range of residue-level interactions ($\sigma \sim 0.6 \text{ nm}$).
To our knowledge, the dimer pair approximation, although intuitive, is novel in that it cannot be reduced to known expansion schemes such as the cluster expansion~\cite{hansenbook} or the expansion proposed by Zimm~\cite{zimm1946application}.
The same approximation should work in other biomolecules, for example, when considering interactions between proteins and RNAs, where coarse-grained simulations have already been conducted~\cite{joseph2021physics,chew2023aromatic}.

Studies on the phase separation behavior of multicomponent systems have used random matrix as the complex interaction between biomolecules~\cite{sear2003instabilities,shrinivas2021phase,zwicker2022evolved}.
Real interactions between biomolecules and especially IDRs should be more restricted compared with a random matrix since the possible interaction types between amino acids are limited~\cite{jacobs2023theory}.
As we have seen, there are only two effective directions that lead to demixing in terms of the difference in the amino acid components, according to the three simulation models that we have tested.
Nevertheless, we found that demixing more than two sequences is possible by appropriately choosing the right directions in the residue difference vector space.
We could not, however, demix more than three distinct sequences in our simulations, consistent with the bound due to the restricted space in the demixing plane.
It will be interesting to see how increasing more components, such as RNAs and phosphorylation of the residues, can rescue the situation in order to explain the nature of the intracellular environment, where there seem to be more than three distinct phases co-existing within just the nucleus.

The prediction of the interactions especially across distinct heteropolymers clearly has room for improvement, as they are still far from perfectly predicting demixing even for the MD simulation results. 
This is likely due to the difficulty in approximating the interaction strength of sequences with large charge blocks, which tended to appear frequently in the demixing candidates.
We also note that our framework is assuming phase separation, and does not predict aggregates~\cite{rana2021phase} in the current form.
Nevertheless, we consider it important to have an analytical model in these calculations as it allows us to construct insights into the rules of demixing. 
In particular, the geometrical argument we proposed in Section \ref{Sec:max_demix} was applicable due to the quadratic form of the effective interaction parameter [Eqs.~\eqref{Eq:B_nondiagonal_vec} and \eqref{Eq:B_nondiagonal_vec_dimer}]. 
It should be useful to consider improved approximation methods that still satisfy this form.

Given the recent intensive studies of protein demixing~\cite{welles2024determinants,rana2024asymmetric}, a crucial step based on our formulation is the experimental validation of the sequences predicted to undergo demixing.
Details on the residue-level interactions in simulation models, however, matter when comparing with the experimental results. 
An interesting approach would be to assess the validity of a microscopic model without relying on MD simulations, which should be beneficial when results of experiments are provided from different conditions including from \textit{in vivo}.
For example, from a large dataset of sequence-to-sequence level interactions in cells, we should be able to restrict and fit the parameters and functional forms of the microscopic model to consistently explain the data by employing optimization methods.

\appendix

\section{Monomer pair approximation}
\label{App:monomer_pair_approximation}

The \secVC between two heteropolymers [Eq.~\eqref{Eq:excluded_volume}] can be expressed as
\begin{equation}
    B (T) = \frac{V}{2} [1 - \braket{e^{- \sum_{n, m} U_{a_n a_m} (|\bm{r}_{1, n} - \bm{r}_{2, m}|) / T}}],
    \label{Eq:excluded_volume_appendix}
\end{equation}
where $V$ is the total volume of the system, $\bm{r}_{1, n}$ ($\bm{r}_{2, m}$) is the coordinate of the $n$th ($m$th) amino acid monomer that constitutes the first (second) polymer, and $a_n$ ($a_m$) is the corresponding amino acid type.
The canonical average $\braket{\cdots}$ is defined as
\begin{equation}
    \braket{\cdots} := \frac{1}{Z} \int \left( \prod_{n, m} d^3 \bm{r}_{1, n} d^3 \bm{r}_{2, m} \right) (\cdots) e^{- [H_\mathrm{intra} (\{ \bm{r}_{1, n} \}) + H_\mathrm{intra} (\{ \bm{r}_{2, m} \})] / T},
\end{equation}
where $Z := \int (\prod_{n, m} d^3 \bm{r}_{1, n} d^3 \bm{r}_{2, m}) e^{- [H_\mathrm{intra} (\{ \bm{r}_{1, n} \}) + H_\mathrm{intra} (\{ \bm{r}_{2, m} \})] / T}$.
Here, $H_\mathrm{intra} (\{ \bm{r}_{1(2), n(m)} \})$ is the intrapolymer Hamiltonian for the first (second) polymer, which consists of the bond interactions between neighboring monomers and the intrapolymer interactions [i.e., $U_{a_n a_{n'}} (|\bm{r}_{1, n} - \bm{r}_{1, n'}|)$ for the first polymer].

We introduce the Mayer $f$-function~\cite{rubinstein2003polymer} between amino acids $a$ and $b$ as
\begin{equation}
    f_{a b} (r) := e^{- U_{a b} (r) / T} - 1.
\end{equation}
Then, we can expand Eq.~\eqref{Eq:excluded_volume_appendix} by the power of $f_{a b}$ as
\begin{align}
    B (T) = & -\frac{V}{2} \sum_{n, m} \braket{f_{a_n a_m} (|\bm{r}_{1, n} - \bm{r}_{2, m}|)} \nonumber \\
    & - \frac{V}{4} \sum_{\substack{n, m, n', m' \\ (n, m) \neq (n', m')}} \braket{f_{a_n a_m} (|\bm{r}_{1, n} - \bm{r}_{2, m}|) f_{a_{n'} a_{m'}} (|\bm{r}_{1, n'} - \bm{r}_{2, m'}|)} \nonumber \\
    & + O(f^3),
    \label{Eq:excluded_volume_expanded_by_f}
\end{align}
where $O(f^3)$ represents the third- and higher-order terms of $f_{ab}$.

The first-order terms in Eq.~\eqref{Eq:excluded_volume_expanded_by_f} are reduced to
\begin{equation}
    B^\mathrm{MP}(T) = - \sum_{n, m} 2 \pi \int_0^\infty dr \, r^2 f_{a_n a_m} (r),
\end{equation}
which is equivalent to Eq.~\eqref{Eq:excluded_volume_monomer_approx}.
Let us introduce the maximum range of the monomer-monomer interaction, $R$.
Given that $f_{a b} (r) \simeq 0$ unless $r < R$ since $U_{a b} (r) \simeq 0$ for $r > R$, $B^\mathrm{MP} (T)$ is on the order of $R^3$.

The monomer pair approximation [$B(T) \simeq B^\mathrm{MP}(T)$ as considered in Sec.~\ref{Sec:boyle_temperature}] should be valid in the following limiting cases.
The first case is when the bond is rigid, and $R$ is much smaller than the bond length $l_b$ (i.e., $R / l_b \ll 1$).
The second case is when the bond is thermally fluctuating with a typical length $l_0$ larger than the natural length, and $R$ is much smaller than $l_0$ (i.e., $R / l_0 \ll 1$).

For the first case, focusing on the second-order terms in Eq.~\eqref{Eq:excluded_volume_expanded_by_f} for $n < n'$ and $m < m'$, we consider the integrals regarding the monomer positions $\{ \bm{r}_{1, n''} \}_{n'' = n}^{n'}$ and $\{ \bm{r}_{2, m''} \}_{m'' = m}^{m'}$ in $\braket{f_{a_n a_m} (|\bm{r}_{1, n} - \bm{r}_{2, m}|) f_{a_{n'} a_{m'}} (|\bm{r}_{1, n'} - \bm{r}_{2, m'}|)}$.
We see that $\braket{f_{a_n a_m} (|\bm{r}_{1, n} - \bm{r}_{2, m}|) f_{a_{n'} a_{m'}} (|\bm{r}_{1, n'} - \bm{r}_{2, m'}|)} \simeq 0$ unless both the pairs $(n, m)$ and $(n', m')$ are spatially close within the distance $R$.
Due to this spatial constraint and the constant bond length, if $\{ \bm{r}_{1, n''} \}_{n'' = n}^{n'}$ and $\bm{r}_{2, m}$ are fixed, the integration by $\{ \bm{r}_{2, m''} \}_{m'' = m + 1}^{m'}$ will involve a factor on the order of $(R / l_b)^2$ or higher, compared to the corresponding integration in partition function $Z$.
After the integration by $\{ \bm{r}_{1, n''} \}_{n'' = n + 1}^{n'}$, the remaining integrals regarding $\{ \bm{r}_{1, n}$, $\bm{r}_{2, m} \}$ will involve a factor on the order of $R^3 / V$ compared to the counterpart in $Z$.
Similar arguments are applied to the cases with $n \geq n'$ or $m \geq m'$.
Overall, the second-order terms in Eq.~\eqref{Eq:excluded_volume_expanded_by_f} should be on the order of $(R / l_b)^2 R^3$.
In the same way, the higher-order terms $O(f^3)$ should be on the order of $(R / l_b)^4 R^3$ or higher.
Thus, in the limit of $R / l_b \to 0$, we obtain $B (T) \to B^\mathrm{MP} (T)$.

For the second case, with the same setup for the monomer coordinates as in the first case, if $\{ \bm{r}_{1, n''} \}_{n'' = n}^{n'}$ and $\bm{r}_{2, m}$ are fixed, the integration by $\{ \bm{r}_{2, m''} \}_{m'' = m + 1}^{m'}$ will involve a factor on the order of $(R / l_0)^3$ or higher, compared to the counterpart in $Z$.
The remaining integrals will involve a factor on the order of $R^3 / V$ compared to the counterpart in $Z$.
Thus, in a similar way to the first case, the second- or higher-order terms in Eq.~\eqref{Eq:excluded_volume_expanded_by_f} should be on the order of $(R / l_0)^3 R^3$ or higher, respectively, leading to $B (T) \to B^\mathrm{MP} (T)$ for $R / l_0 \to 0$.
In the weak bond limit where monomers can freely move as a gas, $l_0$ is regarded as $V^{1/3}$, and $B (T) \to B^\mathrm{MP} (T)$ for large systems with $R / V^{1/3} \ll 1$.

\begin{figure}[t]
    \centering
    \includegraphics[scale=1]{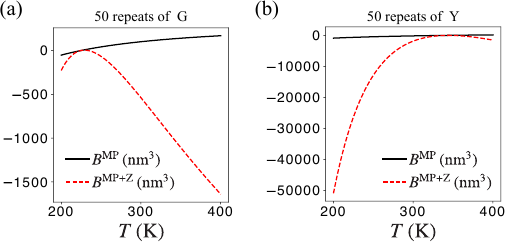}
    \caption{Comparison of \secVCs calculated by the monomer pair approximation and the method proposed in Ref.~\cite{zimm1946application}.
    (a, b) As examples, we plot the temperature dependence of $B^\mathrm{MP}$ (black solid line) and $B^\mathrm{MP + Z}$ (red dashed line) for 50 repeats of (a) G and (b) Y, using the HPS-Tesei model.}
    \label{Fig:zimm}
\end{figure}

\section{Correction to monomer pair approximation}
\label{App:monomer_pair_zimm}
As an approximation scheme to correct the monomer pair approximation (Appendix~\ref{App:monomer_pair_approximation}), we consider the method proposed in Ref.~\cite{zimm1946application} to calculate the second order term in Eq.~\eqref{Eq:excluded_volume_expanded_by_f}.
To use this method, we neglect the intrapolymer interactions, and replace the bond Hamiltonian (with the natural length $l_b$ and the spring constant $k$),
\begin{equation}
    H_\mathrm{bond} := \frac{k}{2} \sum_{n = 1}^{N - 1} (|\bm{r}_{n + 1} - \bm{r}_n| - l_b)^2,
\end{equation}
by
\begin{equation}
    H'_\mathrm{bond} := \frac{3 T}{2 {l_b}^2} \sum_{n = 1}^{N - 1} |\bm{r}_{n + 1} - \bm{r}_n|^2.
\end{equation}
The coefficient of $H'_\mathrm{bond}$ is chosen such that $\braket{|\bm{r}_{n + 1} - \bm{r}_n|^2} = {l_b}^2$ in equilibrium for a single polymer.

Using
\begin{equation}
    \int d^3 \bm{r}_{n + 1} e^{-a |\bm{r}_{n + 1} - \bm{r}_n|^2 - b |\bm{r}_{n + 2} - \bm{r}_{n + 1}|^2} = \frac{\pi^{3 / 2}}{(a + b)^{3 / 2}} e^{-a b |\bm{r}_{n + 2} - \bm{r}_n|^2 / (a + b)}
\end{equation}
repeatedly ($a, b > 0$), we can obtain
\begin{align}
    & \braket{f_{a_n a_m} (|\bm{r}_{1, n} - \bm{r}_{2, m}|) f_{a_{n'} a_{m'}} (|\bm{r}_{1, n'} - \bm{r}_{2, m'}|)} \nonumber \\
    & = \frac{1}{V} \left( \frac{3}{2 \pi {l_b}^2} \right)^3 \frac{1}{|n' - n|^{3 / 2} |m' - m|^{3 /2}} \int d^3 \bar{\bm{r}}_1 d^3 \bar{\bm{r}}_2 d^3 \bar{\bm{r}}_3 \nonumber \\
    & \times f_{a_n a_m} (|\bar{\bm{r}}_2|) f_{a_{n'} a_{m'}} (|\bar{\bm{r}}_3 - \bar{\bm{r}}_1|) \, e^{-3 |\bar{\bm{r}}_1|^2 / (2 {l_b}^2 |n' - n|) - 3 |\bar{\bm{r}}_3 - \bar{\bm{r}}_2|^2 / (2 {l_b}^2 |m' - m|)},
    \label{Eq:second_order_mayer}
\end{align}
where $\bar{\bm{r}}_1 := \bm{r}_{1, n'} - \bm{r}_{1, n}$, $\bar{\bm{r}}_2 := \bm{r}_{2, m} - \bm{r}_{1, n}$, $\bar{\bm{r}}_3 := \bm{r}_{2, m'} - \bm{r}_{1, n}$, and we assume $n \neq n'$ and $m \neq m'$.
Then we replace $\bar{\bm{r}}_3 - \bar{\bm{r}}_2$ by $\bar{\bm{r}}_1$ (i.e., $\bm{r}_{2, m'} - \bm{r}_{2, m}$ by $\bm{r}_{1, n'} - \bm{r}_{1, n}$) in the exponential factor of Eq.~\eqref{Eq:second_order_mayer}, which will be justified when the monomer interaction range is much shorter than $l_b$, similarly to the condition where the monomer pair approximation is valid.
With this replacement, we can perform the integration by $\bar{\bm{r}}_3$ in Eq.~\eqref{Eq:second_order_mayer}, which leads to
\begin{align}
    & \braket{f_{a_n a_m} (|\bm{r}_{1, n} - \bm{r}_{2, m}|) f_{a_{n'} a_{m'}} (|\bm{r}_{1, n'} - \bm{r}_{2, m'}|)} \nonumber \\
    & = \frac{4}{V} \left( \frac{3}{2 \pi {l_b}^2} \right)^{3 / 2} \frac{v_{a_n a_m}^\mathrm{MP} v_{a_{n'} a_{m'}}^\mathrm{MP}}{(|n' - n| + |m' - m|)^{3 /2}},
    \label{Eq:second_order_mayer2}
\end{align}
where $v_{a b}^\mathrm{MP} = -2 \pi \int dr \, r^2 f_{a b} (r)$ [Eq.~\eqref{Eq:excluded_volume_monomer}].

Using Eq.~\eqref{Eq:second_order_mayer2} in the second-order terms of Eq.~\eqref{Eq:excluded_volume_expanded_by_f} and neglecting the $n = n'$ or $m = m'$ terms, we obtain an approximation $B (T) \approx B^\mathrm{MP + Z} (T)$ that corrects the monomer pair approximation as
\begin{equation}
    B^\mathrm{MP + Z} := B^\mathrm{MP} - \left( \frac{3}{2 \pi {l_b}^2} \right)^{3 / 2} \sum_{\substack{n, m, n', m' \\ (n \neq n', m \neq m')}} \frac{v_{a_n a_m}^\mathrm{MP} v_{a_{n'} a_{m'}}^\mathrm{MP}}{(|n' - n| + |m' - m|)^{3 /2}}.
    \label{Eq:zimm_approx}
\end{equation}

In Fig.~\ref{Fig:zimm}, we plot the temperature dependence of $B^\mathrm{MP + Z}$, compared with that of $B^\mathrm{MP}$ for example sequences.
We find that $B^\mathrm{MP + Z}(T)$ shows a non-monotonic temperature dependence and is spuriously negative for high temperatures due to the quadratic contribution of $v^\mathrm{MP}$ in Eq.~\eqref{Eq:zimm_approx}.
This result indicates that Eq.~\eqref{Eq:excluded_volume_expanded_by_f} is not useful as an expansion scheme; we must take in multiple higher-order terms in $f$ in order to recover even the monotonic $T$ dependence of $B(T)$.
We therefore considered the dimer pair approximation as an alternative proxy to estimate the temperature dependence of the \secVC (Sec.~\ref{Sec:improved_prediction}).

\section{Selection of disordered region sequences for simulation}
\label{App:selection_idr_sequences}
We aimed to use disordered region sequences from a wide variety of proteins with identified spatial clustering inside cells. 
To this end, we took data from the Human Cell Map~\cite{go2021proximity} which provides a list of 4,145 human proteins clustered into 20 compartments (`MMF localization') based on proximity labels by biotinylation in HEK293 cells. We took between 2 to 39 proteins from each compartment that have long IDR regions (i.e., pLDDT score from AlphaFold2~\cite{wilson2022alphafold2,jumper2021highly} lower than 0.7 for at least 50 consecutive residues), and selected 270 amino acid sequences of length 50 from these regions.
The list of the selected sequences is presented in the Supplementary Tables~1 (for the HPS-Tesei model) and 2 (for the HPS-Dignon model)~\cite{SM}.

\begin{figure}[t]
    \centering
    \includegraphics[scale=1]{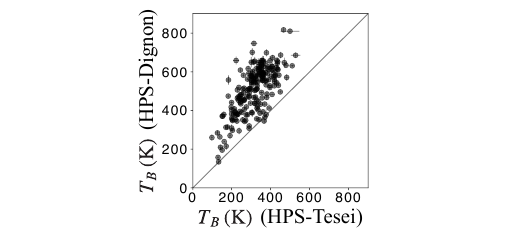}
    \caption{Comparison of the Boyle temperatures between the HPS-Tesei and HPS-Dignon models.
    For 217 IDR sequences, we plot $T_B$ obtained by simulations of each model.
    The gray line is $T_B \, (\text{HPS-Tesei}) = T_B \, (\text{HPS-Dignon})$.}
    \label{Fig:compare_boyle}
\end{figure}

\begin{figure}[t]
    \centering
    \includegraphics[scale=1]{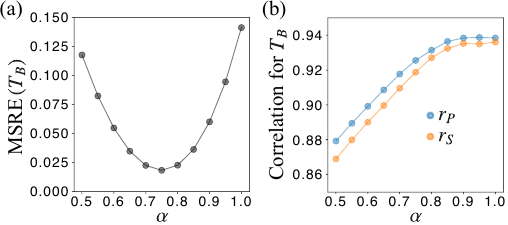}
    \caption{The error and correlation for the Boyle temperature with the rescaled dimer pair approximation.
    (a) The MSRE between $T_B^\mathrm{RDP}$ and $T_B$ as a function of the rescaling parameter $\alpha$.
    (b) The corresponding Pearson correlation coefficient $r_P$ (blue) and Spearman's rank correlation coefficient $r_S$ (orange).}
    \label{Fig:optimize_alpha}
\end{figure}

\section{Selection of $\alpha$}
\label{App:rough_optimization_alpha}

We define the MSRE between $T_B^\mathrm{RDP}$ calculated using Eq.~\eqref{Eq:dimer_interaction_rescaled} and $T_B$ obtained by simulations:
\begin{equation}
    \mathrm{MSRE} \, (T_B) := \frac{1}{N_\mathrm{seq}} \sum_{i = 1}^{N_\mathrm{seq}} \left( \frac{T_{B, i}^{\mathrm{RDP}} - T_{B, i}}{T_{B, i}} \right)^2,
    \label{Eq:msre_boyle}
\end{equation}
where the subscript $i$ of $T_{B, i}^{\mathrm{RDP}}$ and $T_{B, i}$ is the label for the polymer sequence, and $N_\mathrm{seq}$ is the total number of examined sequences.

In Fig.~\ref{Fig:optimize_alpha}(a), we show $\mathrm{MSRE} \, (T_B)$ as a function of the rescaling parameter $\alpha$, which is obtained for $N_\mathrm{seq} = 275$ using the HPS model with the parameter values proposed in Ref.~\cite{tesei2021accurate} (see Sec.~\ref{Sec:prediction_boyle_critical}).
$\mathrm{MSRE} \, (T_B)$ is minimized at $\alpha = 0.75$, which is used in Figs.~\ref{Fig:boyle}(e) and (f).
In Fig.~\ref{Fig:optimize_alpha}(b), we show the $\alpha$ dependence of the Pearson correlation coefficient $r_P$ and Spearman's rank correlation coefficient $r_S$ for the same data set as used in Fig.~\ref{Fig:optimize_alpha}(a).
We find that $r_P$ and $r_S$ are still high ($r_P = 0.926$ and $r_S = 0.919$) when tuning $\alpha$ to $0.75$.

\section{Selection of $l$}
\label{App:rough_optimization_l}

Similar to Eq.~\eqref{Eq:msre_boyle}, we define the MSRE between $T_c^\mathrm{RDP}$ calculated theoretically and $T_c$ obtained by simulations:
\begin{equation}
    \mathrm{MSRE} \, (T_c) := \frac{1}{N_\mathrm{seq}} \sum_{i = 1}^{N_\mathrm{seq}} \left( \frac{T_{c, i}^{\mathrm{RDP}} - T_{c, i}}{T_{c, i}} \right)^2.
    \label{Eq:msre_critical}
\end{equation}

In Fig.~\ref{Fig:critical}(c), we show the $l/l_b$ dependence of $\mathrm{MSRE} \, (T_c)$ as the error of the $T_c$ prediction for each model for the same data set as used in Figs.~\ref{Fig:critical}(a) and (b).
Specifically, we used $\alpha = 0.75$ for all models and took $N_\mathrm{seq} = 233$, $154$, and $73$ for the HPS-Tesei, HPS-Dignon, and Mpipi models, respectively.
For the HPS-Tesei, HPS-Dignon, and Mpipi models, the errors are minimized at $l / l_b$ values of approximately 3.0, 3.2, and 3.6, respectively, based on tests performed at increments of 0.2.

\begin{figure}[t]
    \centering
    \includegraphics[scale=1]{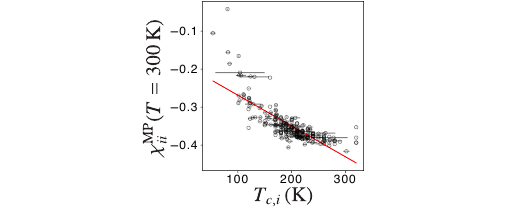}
    \caption{The diagonal interaction parameter $\chi_{ii}^\mathrm{MP}$ at $T=300 \text{ K}$ against $T_{c,i}$ in numerics for the HPS-Tesei model.
    The red line is the result of linear fitting, which we used to convert the distance in Fig.~\ref{Fig:triplet_projection} to $T_c$.
    We took $l/l_b = 3$, to determine the values of $\chi_{ii}^\mathrm{MP}$ in this plot, but the radii of circles plotted in Fig.~\ref{Fig:triplet_projection}(f) do not depend on this choice.}
    \label{Fig:rev_fit_chiMP300K}
\end{figure}

\begin{figure}[t]
    \centering
    \includegraphics[scale=1]{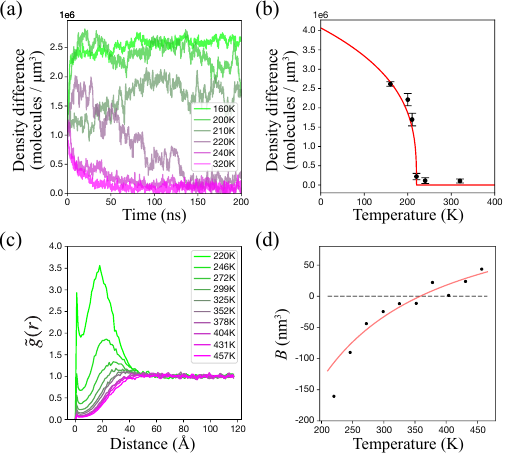}
    \caption{$T_c$ and $T_B$ calculated from MD simulation of the HPS-Tesei model for an example sequence Q9ULK5\_0\_50.
    (a) Density difference between the high-density region and the low-density region as a function of time for simulations at different temperatures.
    (b) Density difference fit with $\widetilde{\Delta \rho}=A(T_c-T)^\beta \theta(T_c-T)$ with $\beta=0.326$ and $\theta(\cdot)$ being the step function.
    (c) $\tilde{g}(r)$ for various temperatures obtained by umbrella sampling and replica exchange.
    (d) $B(T)$ calculated from $\tilde{g}(r)$ according to Eq.~\eqref{Eq:B_from_g}, fit with $B(T)=A_0(1-T/T_B)$.
    }
    \label{Fig:mdsimulation_summary}
\end{figure}

\begin{figure}[t]
    \centering
    \includegraphics[scale=1]{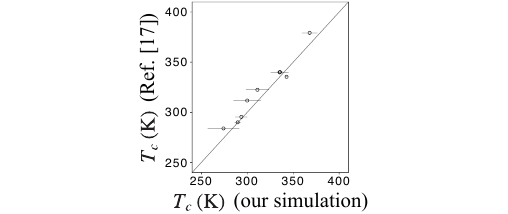}
    \caption{Comparison of $T_c$ for the variants of hnRNPA1 sequences between Ref.~\cite{joseph2021physics} and our simulations.}
    \label{Fig:hnrnp_compare}
\end{figure}

\begin{figure}[t]
    \centering
    \includegraphics[scale=1]{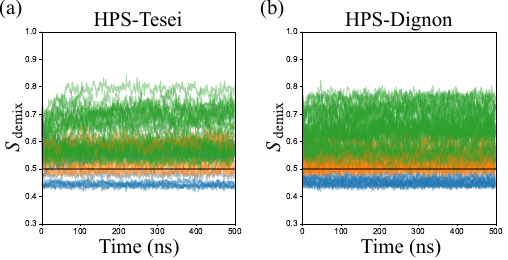}
    \caption{Time evolution of the demixing score $S_\mathrm{demix}$ in MD simulations.
    (a) For different sequences, we plot $S_\mathrm{demix}$ as a function of time for the HPS-Tesei model.
    The colors correspond to $\tilde{\chi}^\mathrm{RDP}_{ij}l^3<-0.04 \text{ nm}^3$ (blue), $-0.04 \text{ nm}^3<\tilde{\chi}^\mathrm{RDP}_{ij}l^3<0.04 \text{ nm}^3$ (orange),  and $\tilde{\chi}^\mathrm{RDP}_{ij}l^3>0.04 \text{ nm}^3$ (green).
    (b) The corresponding plot for the HPS-Dignon model.}
    \label{Fig:demixing_evol}
\end{figure}

\begin{figure*}[t]
    \centering
    \includegraphics[scale=1]{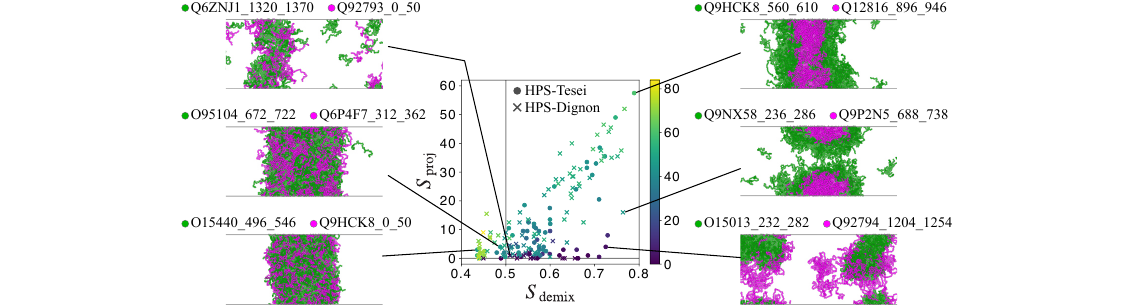}
    \caption{Comparison of the demixing score $S_\mathrm{demix}$ and another indicator of demixing $S_\mathrm{proj}$ (e.g., Ref.~\cite{chew2023thermodynamic}).
    We plot the value for each sequence pair, tested with the HPS-Tesei model (circles) and HPS-Dignon model (crosses).
    The brightness indicates the number density at the angular mean position, which should be high for phase-separated condensates and low when there are multiple regions of high-density regions, which is typical in the presence of aggregate-like structures.
    We also show typical configurations of the sequence pair mixtures.}
    \label{Fig:rev_demix}
\end{figure*}

\begin{figure}[t]
    \centering
    \includegraphics[scale=1]{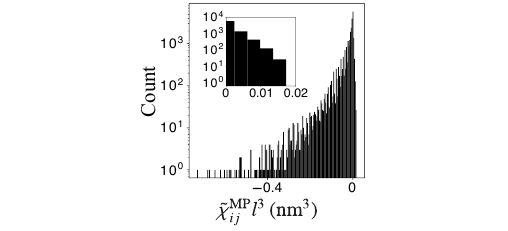}
    \caption{Histogram of the effective inter-component interaction parameter calculated by the monomer pair approximation, $\tilde{\chi}_{ij}^\mathrm{MP}$, at $T = \min \{ T_{c, i}, T_{c, j} \}$.
    The inset is the enlarged view of the positive region of $\tilde{\chi}_{ij}^\mathrm{MP}$.
    See Fig.~\ref{Fig:demixing} for the counterpart when using the rescaled dimer pair approximation.}
    \label{Fig:demixing_dist_supplement}
\end{figure}

\begin{figure}[t]
    \centering
    \includegraphics[scale=1]{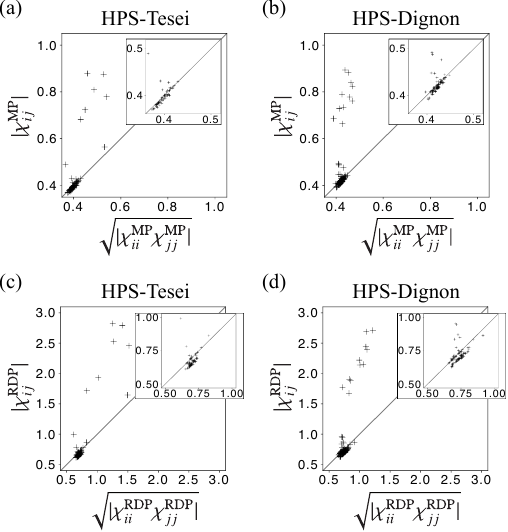}
    \caption{Comparison between the off-diagonal interaction parameter and the geometric mean of the diagonal interaction parameters.
    (a, b) We compare $|\chi_{ij}^\mathrm{MP}|$ and $\sqrt{|\chi_{ii}^\mathrm{MP} \chi_{jj}^\mathrm{MP}|}$ at $T = \min \{ T_{c, i}, T_{c, j} \}$ obtained by the monomer pair approximation for the (a) HPS-Tesei and (b) HPS-Dignon models.
    The gray line is $|\chi_{ij}^\mathrm{MP}| = \sqrt{|\chi_{ii}^\mathrm{MP} \chi_{jj}^\mathrm{MP}|}$, and each inset is an enlarged view.
    (c, d) The corresponding plots with the rescaled dimer pair approximation.
    We took $l/l_b = 3$ for all the results here.}
    \label{Fig:compare_rpa}
\end{figure}

\begin{figure}[t]
    \centering
    \includegraphics[scale=1]{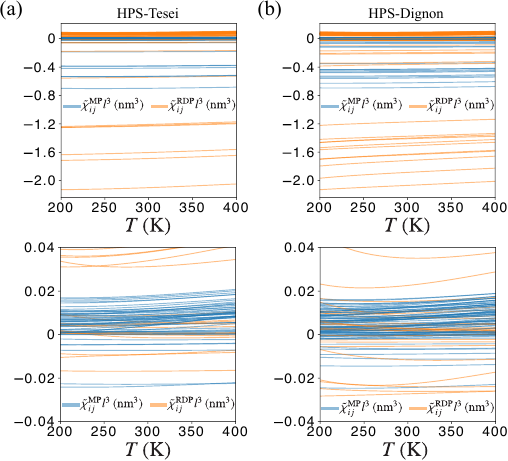}
    \caption{Temperature dependence of the effective inter-component interaction parameters calculated by the monomer and rescaled dimer pair approximations.
    (a) For $62$ sequence pairs used for Fig.~\ref{Fig:demixing}(c), we plot the temperature dependence of $\tilde{\chi}_{ij}^\mathrm{MP} (T)$ and $\tilde{\chi}_{ij}^\mathrm{RDP} (T)$ obtained for the HPS-Tesei model, with the overall (upper panel) and enlarged (lower panel) views.
    (b) For $94$ sequence pairs used for Fig.~\ref{Fig:demixing}(c), we plot the counterpart of (a) for the HPS-Dignon model.}
    \label{Fig:Btilde_Tdep}
\end{figure}

\begin{figure}[t]
    \centering
    \includegraphics[scale=1]{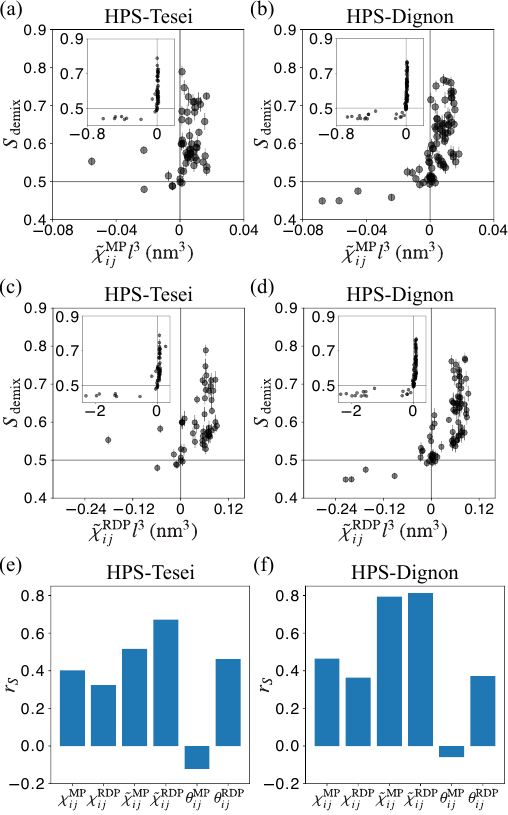}
    \caption{Comparison between the demixing score $S_\mathrm{demix}$ and the off-diagonal interaction parameter.
    (a, b) We compare $S_\mathrm{demix}$ and $\tilde{\chi}_{ij}^\mathrm{MP}$ at $T = \min \{ T_{c, i}, T_{c, j} \}$ obtained by the monomer pair approximation for the (a) HPS-Tesei and (b) HPS-Dignon models.
    The gray horizontal and vertical lines are $S_\mathrm{demix} = 0.5$ and $\tilde{\chi}_{ij}^\mathrm{MP} = 0$, respectively, and each inset is a reduced view that covers a wider range in the horizontal axis.
    (c, d) The corresponding plots with the rescaled dimer pair approximation.
    (e, f) Spearman's rank correlation coefficient $r_S$ between $S_\mathrm{demix}$ and each quantity defined in Sec.~\ref{SubSec:predict_demixing_two_comp} for the (e) HPS-Tesei and (f) HPS-Dignon models.}
    \label{Fig:demixing_supplement}
\end{figure}

\begin{figure}[t]
    \centering
    \includegraphics[scale=1]{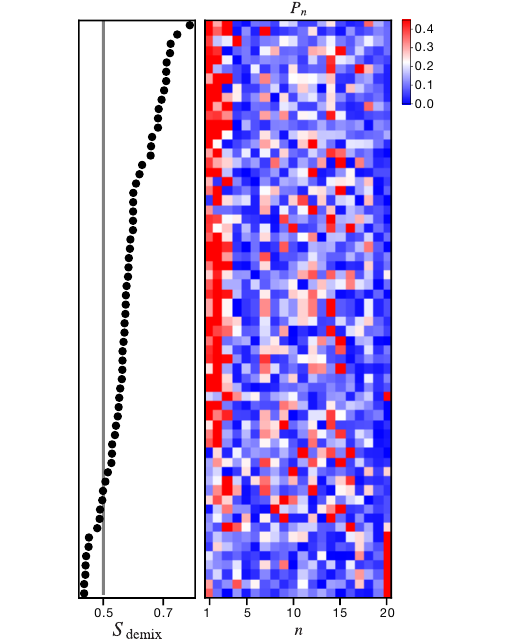}
    \caption{Full result on the eigendecomposition of the amino acid fraction difference vector $\bm{n}_j - \bm{n}_i$ for the pairs tested in the two-component simulation (Fig.~\ref{Fig:demixing}).}
    \label{Fig:projection_all}
\end{figure}

\begin{figure}[t]
    \centering
    \includegraphics[scale=1]{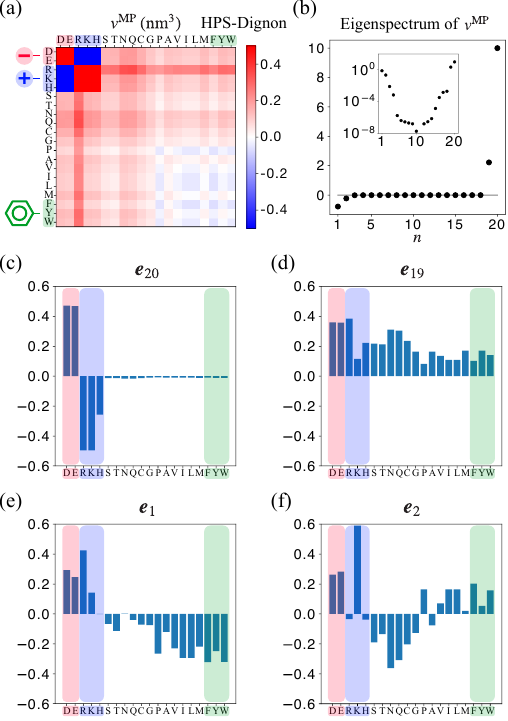}
    \caption{(a) \secVC matrix calculated at $300\text{ K}$ in the monomer pair approximation using the HPS-Dignon parameters. (b) Eigenspectrum of the \secVC matrix. Inset shows the absolute values of the eigenvalues with logscale. (c) Eigenvector with the largest eigenvalue. (d) Eigenvector with the second largest eigenvalue. (e) Eigenvector with the most negative eigenvalue. (f) Eigenvector with the second most negative eigenvalue. For all the plots of eigenvectors, the component for amino acid D is taken as positive.}
    \label{Fig:exvol_matrix_dignon}
\end{figure}

\begin{figure}[t]
    \centering
    \includegraphics[scale=1]{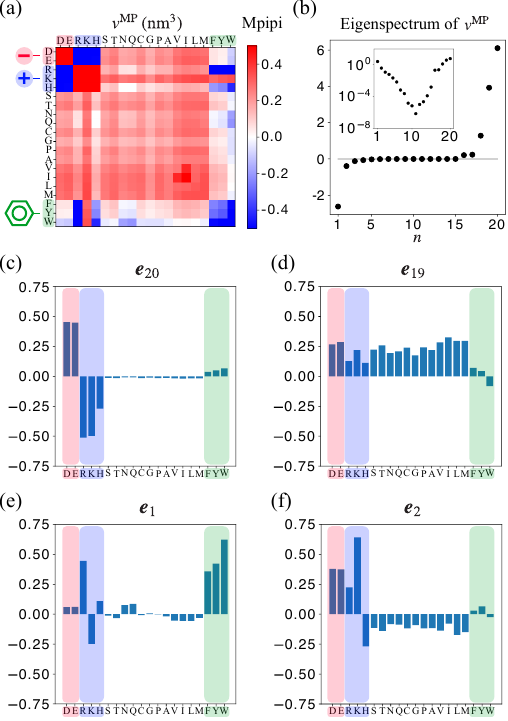}
    \caption{(a) \secVC matrix calculated at $300\text{ K}$ in the monomer pair approximation using the Mpipi parameters. (b) Eigenspectrum of the \secVC matrix. Inset shows the absolute values of the eigenvalues with logscale. (c) Eigenvector with the largest eigenvalue. (d) Eigenvector with the second largest eigenvalue. (e) Eigenvector with the most negative eigenvalue. (f) Eigenvector with the second most negative eigenvalue. For all the plots of eigenvectors, the component for amino acid D is taken as positive.}
    \label{Fig:exvol_matrix_mpipi}
\end{figure}

\begin{figure*}[t]
    \centering
    \includegraphics[scale=1]{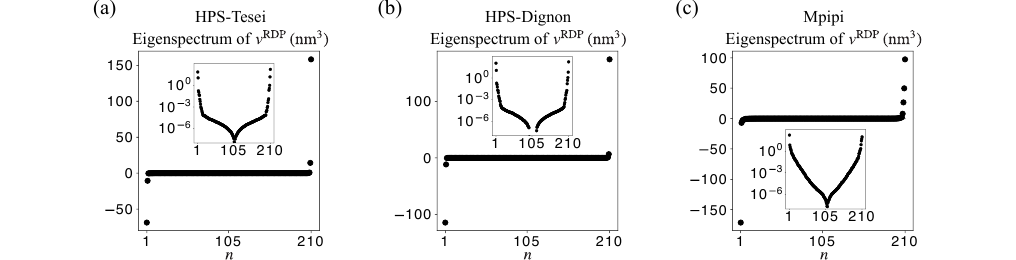}
    \caption{Eigenspectrum of $v^\mathrm{RDP} (T = 300 \text{ K})$, the \secVC matrix calculated by the rescaled dimer pair approximation, for the (a) HPS-Tesei, (b) HPS-Dignon, and (c) Mpipi models.
    Each inset shows the absolute values of the eigenvalues with logscale.
    We omitted $190$ zero eigenvalues that are derived from $v^\mathrm{RDP}_{d \bar{d}} = v^\mathrm{RDP}_{\bar{d} d}$ for $d \neq \bar{d}$, where the dimer $\bar{d}$ is the reverse of $d$ [e.g., $\bar{d} = (\mathrm{F}, \mathrm{A})$ for $d = (\mathrm{A}, \mathrm{F})$].}
    \label{Fig:exvol_eigval_dimer}
\end{figure*}

\section{Estimation of the relation between $T_{c,i}$ and $\chi_{ii}^\mathrm{MP}$}
\label{App:estimate_critical_temp_from_chimp}

To connect the value of $\chi_{ii}^\mathrm{MP}$ at $T=300 \text{ K}$ with $T_{c,i}$ in numerics, we calculated $\chi_{ii}^\mathrm{MP}$ in the HPS-Tesei model for the same sequence set as used in Fig.~\ref{Fig:critical}.
In Fig.~\ref{Fig:rev_fit_chiMP300K}, we plot $\chi_{ii}^\mathrm{MP}$ against $T_{c,i}$ from simulations (black dots), showing negative correlation ($r_P = -0.78$ and $r_S = -0.82$).
By linear fitting (red line), we calculated the typical value of $\chi_{ii}^\mathrm{MP}$ for a given $T_{c,i}$.
For $T_{c,i} = 210\text{ K}$, $220\text{ K}$, $230\text{ K}$, and $240\text{ K}$, we obtained $\chi_{ii}^\mathrm{MP}$, which is transformed into circles in the demixing plane [Fig.~\ref{Fig:triplet_projection}(f)] using Eq.~\eqref{Eq:B_diagonal_vec_approx} with fixed $\tilde{n}_{i, 19} = 0.46 \text{ nm}^{3/2}$ and $\tilde{n}_{i, 20} = 0$.

\section{Simulation}
We used GENESIS 2.0.0~\cite{tan2022implementation,jung2015genesis,kobayashi2017genesis} to conduct MD simulations of the coarse-grained polypeptide chains. We used the Langevin thermostat with a friction coefficient of $0.01\text{ ps}^{-1}$, with $0.01\text{ ps}$ time steps.
The parameters $\{ q_a \}$, $\{ \sigma_a \}$, and $\{ \lambda_a \}$ for the HPS models that we used are presented in Supplementary Tables~3 (for the HPS-Tesei model) and 4 (for the HPS-Dignon model)~\cite{SM}.
The parameters for the Mpipi model we used are presented in Supplementary Tables~5 and 6~\cite{SM}
For simulation as well as in theoretical calculations, we took $\epsilon = 0.8368 \text{ kJ/mol}$, $\varepsilon_r = (249.4 - 0.788 T + 7.2 \times 10^{-4} T^2) (1 - 0.2551 c_s + 5.151 \times 10^{-2} {c_s}^2 - 6.889 \times 10^{-3} {c_s}^3)$ for $T \, \text{(K)}$ and $c_s \, \text{(mol/l)}$, following Ref.~\cite{tan2022implementation}, and set $c_s = 0.15 \text{ mol/l}$.
For visualization, we used napari~\cite{napari2019} in Python.

For the Mpipi simulations, we used the LAMMPS code provided in \cite{joseph2021physics} but with 65 distinct IDR sequences taken from the IDRome~\cite{tesei2024conformational} that have the length of 135 residues.

\subsection{Critical temperature}
\label{App:simulation_critical_temperature}
To predict the critical temperature of phase separation, $T_c$, we conducted numerical simulations at various temperatures.

\subsubsection{HPS-Tesei and HPS-Dignon model simulation using GENESIS}

For the HPS-Tesei and HPS-Dignon model simulations, we used the periodic boundary condition with $18\text{ nm} \times 18\text{ nm} \times 200\text{ nm}$ containing $N$ molecules.
As the initial condition, we first conducted a simulation of a single chain at $150\text{ K}$ for $10\text{ ns}$ ($10^6\text{ steps}$), and copied that configuration $N$ times within a small region inside the box, utilizing duplication\_generator.jl provided in GENESIS.

We conducted a binary search scheme at $10\text{ K}$ resolution between $0\text{ K}$ and $630\text{ K}$ to run the simulations at appropriate temperatures.
We first conducted simulations at $T_0=320\text{ K}$ for $200\text{ ns}$ ($2\times 10^7\text{ steps}$). 
The density profile was obtained by first projecting the distribution of the molecules to the z-axis.
The density of the high-density region $\rho_H(t)$ was calculated as the peak value of the z-profile, and the density of the low-density region $\rho_L(t)$ as the density at the position 100 nm away from the position that was used to calculate $\rho_H(t)$. Both of these densities were calculated for the whole time course for each timepoint $t$.
Looking at $\Delta \rho (t):=\rho_H(t)-\rho_L(t)$ and its mean value for a set time interval $\widetilde{\Delta \rho} (t_0,t_1):=\sum_{t_0 \leq t<t_1} \Delta \rho (t)/(t_1-t_0)$, we decided to increase the temperature in the next simulation when 
\begin{eqnarray}
    \frac{\widetilde{\Delta \rho} (t_m,t_e)}{\widetilde{\Delta \rho} (0,t_i)} > 0.7 \ \ \& \ \  \min_{t \geq t_m} \Delta \rho (t) > 100, \label{Eq:updowncondition}
\end{eqnarray}
and decreased the temperature otherwise. 
The condition \eqref{Eq:updowncondition} was set empirically with $t_i=10\text{ ns}$ and $t_m=150\text{ ns}$ to capture whether the equilibrated state is phase-separated or not within the finite time of the simulation.
After deciding to raise (or lower) the temperature for the simulation, we selected the midpoint between the most recently simulated temperature and the nearest larger (or smaller) value with a maximum of 630 K (or the minimum of 10 K), and simulated another $200\text{ ns}$ ($2\times 10^7\text{ steps}$)after re-preparing the initial condition described above.
Following this procedure, the simulation will stop after running for six distinct temperatures [e.g.,  $320\text{ K}$, $160\text{ K}$, $240\text{ K}$, $200\text{ K}$, $220\text{ K}$, and $210\text{ K}$ in the case of Q9ULK5\_0\_50, as shown in Fig.~\ref{Fig:mdsimulation_summary}(a)].

After gathering all the simulation results for the six rounds, we fit the density difference to 
\begin{eqnarray}
\widetilde{\Delta \rho} (t_m,t_e) = A(T_c - T)^\beta \label{Eq:rho_T_fit}
\end{eqnarray}
with $\beta=0.326$ using the recent result of the three-dimensional Ising critical exponent~\cite{simmons2015semidefinite} [see Fig.~\ref{Fig:mdsimulation_summary}(b) for the example case of Q9ULK5\_0\_50].
We obtained the goodness of fit measured by the reduced $\chi^2$  and the monotonicity of the plot by calculating Spearman's rank correlation coefficient ($r_S$) between $\widetilde{\Delta \rho} (t_m,t_e)$ and $T$.
On top of the filtering we applied in calculating $T_B$ (see next section), we filtered out the data of the sequences that had reduced $\chi^2 > 20$ or the error of fit in $T_c$ that are larger than $50\text{ K}$ considering that the $T_c$ fits cannot be reliable in those cases.

\subsubsection{Mpipi model simulation using LAMMPS}

For the Mpipi model simulations, we used the initial condition profile as well as the parameters for the time evolution as provided in the code from \cite{joseph2021physics}. The periodic boundary condition was set as $10\text{ nm} \times 10\text{ nm} \times 44\text{ nm}$ containing $63$ molecules, and the simulations were run up to $40\text{ ns}$ after $5\text{ ns}$ of equilibration.
To calculate $T_c$, we first estimated its value using the rescaled dimer-pair approximation, using $l/l_b=3.0$.
Then we took up to 12 points around this value with at least 10 K intervals between them and calculated the projected density in the $z$-axis after the simulation to obtain the density difference, and fit using Eq.~\eqref{Eq:rho_T_fit}   as in the case of the HPS-Tesei and HPS-Dignon model simulations. We checked the validity of this approach by comparing it to the values of $T_c$ provided in \cite{joseph2021physics} for the variants of hnRNPA1 sequences (Fig.~\ref{Fig:hnrnp_compare}).

\subsection{Calculating $B(T)$}
\label{App:simulation_excluded_volume}
We employed umbrella sampling with exchange Monte Carlo simulation to obtain $B(T)$ using two chains of the same species.
To determine the range of temperatures to conduct the simulation, we assumed that $T_B$ should be roughly between $T_c$ and $2T_c$, and selected  $T = (1+0.12 n ) \tilde{T}_c $ for $n=0,1,...,10$.

For the exchange Monte Carlo simulation with umbrella sampling, we applied the harmonic biasing potential to the center of mass distance between the two proteins with a spring constant of 0.1 kcal/(mol \AA $^2$). 
We chose 20 values for $d_0$, the center of the distance in the umbrella sampling, as $d_0=
0,\ 4,\ 8,...,\ 60,\ 68,\ 76,\ 84,\ 92,\ 100 \ \text{\AA}$.
The simulation was conducted in $25\  \text{nm} \times 25\  \text{nm} \times 25\text{ nm}$ periodic boundary boxes, with the exchange period set as $1\text{ ns}$ ($10^5\text{ steps}$) and the total run of $100\text{ ns}$ ($10^7\text{ steps}$).
Trajectory extraction and the weighted histogram analysis method (WHAM) were conducted using the pipelines in GENESIS~\cite{tan2022implementation}.
From the obtained potential of mean force (PMF) between $0 \ \text{\AA} \leq r < N_\mathrm{max} \ \text{\AA}$ with $N_\mathrm{max}=118$ at $\delta r=1$ \AA{} resolution, we calculated $B(T)$ by the integral up to 70 \AA:
\begin{equation}
    B(T)=2 \pi \delta r \sum_{0 \leq i < N_B} [1-\tilde{g}(r_i)] r_i^2, \label{Eq:B_from_g}
\end{equation}
with $r_i=i\delta r$ and $N_B=70$. 
For $\tilde{g}(r)$, we first obtained ${g}(r)$ from the PMF using the temperature of each simulation and normalized it as $\tilde{g}(r):={g}(r)/\sum_{N_B \leq i < N_\mathrm{max}} {g}(r_i)/(N_\mathrm{max}-N_B)$ [see Fig.~\ref{Fig:mdsimulation_summary}(c) for the example case of Q9ULK5\_0\_50].

We numerically obtained the Boyle temperature $T_B$ by fitting $B(T)$ to $A_0(1-T/T_B)$ [see Fig.~\ref{Fig:mdsimulation_summary}(d) for the example case of Q9ULK5\_0\_50].
To see the goodness of fit, we calculated the coefficient of determination ($R^2$) and the error for $T_B$ due to the curve fitting. We excluded the data for sequences where $R^2$ was smaller than $0.6$ or the error of $T_B$ was larger than $50\text{ K}$ considering that those values are unreliable.

We provide the result of the $T_B$ and $T_c$ obtained from the MD simulation as Supplementary Tables~1 (for the HPS-Tesei model) and 2 (for the HPS-Dignon model)~\cite{SM}.

\subsection{Multi-component demixing}
\label{App:multicomponent}
We conducted the $M$-component simulations by using the same boundary condition and similar slab initial condition for the $T_c$ estimation but with $200$ molecules each.

For the two-component simulations, we took the pairs of sequences with a difference in $T_c$ (estimated from simulation) smaller than $60\text{ K}$, and selected the pairs that had the largest ($>0.04 \text{ nm}^3$), mid-ranged (between $-0.04 \text{ nm}^3$ and $0.04 \text{ nm}^3$), and negative values ($<-0.04 \text{ nm}^3$) of $\tilde{\chi}_{ij}(T) l^3$. Simulations were conducted for $500 \text{ ns}$ for each pair.
For the temperature in the simulation, we chose $T = \left\lfloor \min \{ T_{c, i}, T_{c, j} \}/10 \right\rfloor \times 10 \  \mathrm{K}$, where $T_{c, i}$ is the critical temperature obtained by single-component MD simulations for a sequence $i$.
We chose this sequence-dependent temperature setting for the multi-component simulation to avoid the potential artifact that can arise when fixing a temperature; since the range of critical temperatures is wide, setting a fixed temperature across simulations will cause certain sequences to be deep in the phase-separated regime whereas other sequences to be marginal or even non-phase separating.

For the demixing matrix $\bm{S}$, we first calculated the mean position of each polymer (indexed by $n,n'$) and constructed a $k$-nearest neighbor ($k$-NN) graph with $k=8$ and a cutoff distance $3$ nm while taking into account the periodic boundary condition.
We then symmetrize the graph by taking the union of $k$-NN graph, where we denote the obtained adjacency (symmetric) matrix as $\bm{s}$ with components $s_{n n'}$. The components of the demixing matrix $(1 \leq i,j \leq M)$ is then
\begin{equation}
    S_{ij} = \frac{\sum_{n \in i} \sum_{n' \in j} s_{n n'}}{\sum_{n,n'} s_{n n'}}.
\end{equation}
where $n \in i$ means that molecule $n$ is species $i$.
The demixing score $S_\mathrm{demix}:=\mathrm{Tr} \, \bm{S}$ used to quantify the extent of demixing in the $M=2$ component case is $S_\mathrm{demix}=S_{11}+S_{22}$, which is the fraction of edges in the union symmetrized $k$-NN graph that connects the same species of molecules over the total edges.

For the case of $M=2$, we compared the demixing score $S_\mathrm{demix}$ with $S_\mathrm{proj}:= |\rho_{i,\mathrm{center}}-\rho_{j,\mathrm{center}}|$ (see Fig.~\ref{Fig:rev_demix}), where $\rho_{i,\mathrm{center}}$ is the number density of component $i$ at the center of the condensate. The center of the condensate was calculated as the angular mean position of the density profile, where we used the sum of the $z$-projected number densities, $\rho_{i}+\rho_{j}$.

We show in Fig.~\ref{Fig:demixing_evol} how $S_\mathrm{demix}$ equilibrates quickly within around $100 \text{ ns}$ in simulation.
The values used in Figs.~\ref{Fig:demixing} and \ref{Fig:demixing_antagonistic} for the two-component simulations are the time average between $300 \text{ ns}$ and $500 \text{ ns}$.
For Fig.~\ref{Fig:demixing}(c), we used the data of sequence pairs with the mean polymer-polymer distance for either type of sequence shorter than $2.65 \text{ nm}$ to exclude the case with no or weak phase separation.

For the $M=3$ component simulation with both HPS-Tesei and HPS-Dignon, we selected the sets of three sequences that either have all the pairs satisfying $\tilde{\chi}^\mathrm{MP}_{ij} (\min\{T_{c}^i,T_{c}^j\})> 0.0044 \text{ nm}^3/l^3$, or all the pairs satisfying $\tilde{\chi}^\mathrm{RDP}_{ij} (\min\{T_{c}^i,T_{c}^j\})> 0.018 \text{ nm}^3/l^3$.
These positive lower bounds on the effective interaction parameters were set to narrow down the candidate sets to the order of tens, and finally selected the top eight candidate triplets in terms of the values of $\tilde{\chi}^\mathrm{MP}_{ij} (\min\{T_{c}^i,T_{c}^j\})l^3$ and $\tilde{\chi}^\mathrm{RDP}_{ij} (\min\{T_{c}^i,T_{c}^j\})l^3$ for both HPS-Tesei and HPS-Dignon, resulting in 32 sets in total.
The simulations were conducted for at least $150 \text{ ns}$ and up to $500 \text{ ns}$ at the temperature corresponding to the lowest critical temperature in the set, and the demixing matrix $\bm{S}$ was calculated as the time average of simulations after $150 \text{ ns}$.

For the $M=4$ component simulation with HPS-Tesei, we selected the sets of four sequences that either have all the pairs satisfying $\tilde{\chi}^\mathrm{MP}_{ij} (\min\{T_{c}^i,T_{c}^j\})> 0.0022 \text{ nm}^3/l^3$, or all the pairs satisfying $\tilde{\chi}^\mathrm{RDP}_{ij} (\min\{T_{c}^i,T_{c}^j\})> 0.0112 \text{ nm}^3/l^3$. For HPS-Dignon, we selected the sets of four sequences that either have all the pairs contained satisfying $\tilde{\chi}^\mathrm{MP}_{ij} (\min\{T_{c}^i,T_{c}^j\})> 0.0024 \text{ nm}^3/l^3$ or all the pairs contained satisfying $\tilde{\chi}^\mathrm{RDP}_{ij} (\min\{T_{c}^i,T_{c}^j\})> 0.014 \text{ nm}^3/l^3$.
These positive lower bounds on the effective interaction prameters were set to narrow down the candidate sets to the order of five to ten, and finally selected the top four candidate triplets in terms of the values of $\tilde{\chi}^\mathrm{MP}_{ij} (\min\{T_{c}^i,T_{c}^j\})l^3$ and $\tilde{\chi}^\mathrm{RDP}_{ij} (\min\{T_{c}^i,T_{c}^j\})l^3$ for both HPS-Tesei and HPS-Dignon, resulting in 16 sets in total.
The simulations were conducted for  $500 \text{ ns}$ at the temperature corresponding to the lowest critical temperature in the set, and the demixing matrix $\bm{S}$ was calculated as the time average of simulations after $300 \text{ ns}$.

\section{Projection and restriction of the sequence vectors}
\label{App:projection}
To obtain the region for $(\tilde{n}_{i, 1}, \tilde{n}_{i, 2})$ where $\tilde{n}_{i, 19}$ and $\tilde{n}_{i, 20}$ are constants ($a$ and $b$, respectively) [bold line region in Fig.~\ref{Fig:triplet_projection}(d)], we first calculated the convex hull formed by 20 points $\{ (\tilde{e}_{a,1}, \tilde{e}_{a,2}, \tilde{e}_{a,19}, \tilde{e}_{a,20}) \}_a$ in the four-dimensional space.
In the following, we explain the remaining procedure to find the cross-section of this convex hull and the three-dimensional hyperplanes, $\tilde{n}_{i, 19} = a$ and $\tilde{n}_{i, 20} = b$.
For simplicity, we rewrite the four-dimensional coordinate $(\tilde{n}_{i, 1}, \tilde{n}_{i, 2}, \tilde{n}_{i, 19}, \tilde{n}_{i, 20})$ as $(x, y, z, w)$.

The obtained convex hull $K$ is a four-dimensional polytope with three-dimensional faces that are represented by triangulation as a union of three-dimensional simplices (i.e., tetrahedra) $\{ S^l \}_l$, each of which consists of four vertices $\{ \bm{v}^{l, 1}, \bm{v}^{l, 2}, \bm{v}^{l, 3}, \bm{v}^{l, 4} \}$.
Each vertex is a point in the four-dimensional space, e.g., $\bm{v}^{l, 1} = (v^{l, 1}_x, v^{l, 1}_y, v^{l, 1}_z, v^{l, 1}_w)$.
There are four two-dimensional faces (i.e., triangles) $\{ F^{l, m} \}_{m = 1}^4$ of each simplex $S^l$, and a point on the face $F^{l, m}$ can be expressed as $(1 - s^{l, m} - t^{l, m}) \bm{v}^{l, m} + s^{l, m} \bm{v}^{l, m + 1} + t^{l, m} \bm{v}^{l, m + 2}$, where $\bm{v}^{l, 5} := \bm{v}^{l, 1}$, $\bm{v}^{l, 6} := \bm{v}^{l, 2}$, and $(s^{l, m}, t^{l, m})$ is the barycentric coordinate that should satisfy $s^{l, m}, t^{l, m} \geq 0$ and $s^{l, m} + t^{l, m} \leq 1$.

To calculate the coordinate of a possible intersection of the face $F^{l, m}$ and the two hyperplanes, $z = a$ and $w = b$, we solve $(x^{l, m}, y^{l, m}, a, b) = (1 - s^{l, m} - t^{l, m}) \bm{v}^{l, m} + s^{l, m} \bm{v}^{l, m + 1} + t^{l, m} \bm{v}^{l, m + 2}$ for the four variables $\{ x^{l, m}, y^{l, m}, s^{l, m}, t^{l, m} \}$.
We can explicitly solve the equations as
\begin{equation}
    \begin{pmatrix}
        s^{l, m} \\
        t^{l, m}
    \end{pmatrix}
    =
    \begin{pmatrix}
        v^{l, m + 1}_z - v^{l, m}_z & v^{l, m + 2}_z - v^{l, m}_z \\
        v^{l, m + 1}_w - v^{l, m}_w & v^{l, m + 2}_w - v^{l, m}_w
    \end{pmatrix}^{-1}
    \begin{pmatrix}
        a - v^{l, m}_z \\
        b - v^{l, m}_w
    \end{pmatrix},
\end{equation}
and then
\begin{equation}
    \begin{pmatrix}
        x^{l, m} \\
        y^{l, m}
    \end{pmatrix}
    =
    \begin{pmatrix}
        (1 - s^{l, m} - t^{l, m}) v^{l, m}_x + s^{l, m} v^{l, m + 1}_x + t^{l, m} v^{l, m + 2}_x \\
        (1 - s^{l, m} - t^{l, m}) v^{l, m}_y + s^{l, m} v^{l, m + 1}_y + t^{l, m} v^{l, m + 2}_y
    \end{pmatrix}.
\end{equation}
If and only if the solution satisfies $s^{l, m}, t^{l, m} \geq 0$ and $s^{l, m} + t^{l, m} \leq 1$, the face $F^{l, m}$ has the intersection $\bm{u}^{l, m} := (x^{l, m}, y^{l, m}, a, b)$.
Repeating this procedure for all $l$ and $m$, we can obtain all the intersections, $\{ \bm{u}^{l, m} \}_{l, m}$.
Finally, we can obtain the cross-section of the convex hull $K$ and the hyperplanes $z = a$ and $w = b$ as the convex hull formed by $\{ \bm{u}^{l, m} \}_{l, m}$.
For the numerical calculation of the convex hull, we used a Python package (scipy.spatial.ConvexHull~\cite{virtanen2020scipy}).

\section{Code availability}
See \url{https://github.com/adachi24/virialcoeff} for the functions and example code to calculate the estimates of virial coefficients, Boyle temperatures, and critical temperatures for a given sequence.

\begin{acknowledgments}
We thank Tomoya Hayata and Takuya Nomoto for their helpful comments. We also acknowledge Cheng Tan and Yuji Sugita for their helpful comments on the simulation methods.
This work was supported by JSPS KAKENHI Grant Numbers JP20K14435 (to K.A.), JP19H05795,	JP19H05275, JP21H01007, and JP23H00095 (to K.K.).
\end{acknowledgments}

\bibliography{ref.bib}

\end{document}